%% file: main.tex
\makeatletter
\def\cl@chapter{}
\makeatother

\documentclass[twocolumn]{svjour3}
\usepackage{etoolbox}
\newtoggle{TR}
\toggletrue{TR}

\iftoggle{TR}{\def\makeheadbox{\relax}}{}

\input{preamble/packages}
\input{preamble/macros}

\begin{document}
\input{contents/title}
\input{contents/abstract}
\input{contents/introduction}
\input{contents/probmodelcheck}

\input{contents/stormfeatures}

\input{contents/stormusage}

\input{contents/evaluation}
\input{contents/conclusion}

\input{contents/acknowledgements}
\bibliographystyle{spmpsci}
\bibliography{literature}
\end{document}

%% file: preamble/packages.tex
\usepackage[T1]{fontenc}
\usepackage{amsmath,amssymb,amsfonts}
\usepackage{xspace}
\usepackage{booktabs}
\usepackage{color}
\usepackage{xcolor}
\usepackage{graphicx}
\usepackage{microtype}
\usepackage{wrapfig}
\usepackage{setspace}
\usepackage{hyperref}
\usepackage{pgfplots}
\usepackage{subfigure}
\usepackage{xifthen}
\usepackage{listings}

\usepackage{tikz}
\usetikzlibrary{fit,petri,arrows,shapes,calc,automata,decorations.pathreplacing,backgrounds,positioning}
\newcommand{\tikzhorizdist}{1.4cm}

\usepackage{pgfplots}
\pgfplotsset{compat=newest}

\usepackage{hyperref}
\hypersetup{}

\usepackage[capitalise,noabbrev]{cleveref}
\crefname{inner_example}{Example}{Examples}
\Crefname{inner_example}{Example}{Examples}

\usepackage{acro}
\acsetup{messages={silent}}
\input{contents/acronyms}

\usepackage[inline]{enumitem}
\setlist[enumerate,1]{label=(\roman*)}

%% file: contents/acronyms.tex
\DeclareAcronym{dd}{
  short=DD,
  long=decision diagram,
}
\DeclareAcronym{bdd}{
  short=BDD,
  long=binary decision diagram,
}
\DeclareAcronym{mtbdd}{
  short=MTBDD,
  long=multi-terminal binary decision diagram,
}
\DeclareAcronym{mdd}{
  short=MDD,
  long=multi-valued decision diagram,
}
\DeclareAcronym{odd}{
  short=ODD,
  long=offset-labeled binary decision diagram,
}

\DeclareAcronym{lts}{
  short=LTS,
  long=labelled transition system,
}
\DeclareAcronym{mc}{
  short=MC,
  long=Markov chain,
}
\DeclareAcronym{dtmc}{
  short=DTMC,
  long=discrete-time Markov chain,
}
\DeclareAcronym{ctmc}{
  short=CTMC,
  long=continuous-time Markov chain,
}
\DeclareAcronym{pa}{
  short=PA,
  long=probabilistic automaton,
  short-plural-form=PA,
  long-plural-form=probabilistic automata
}
\DeclareAcronym{mdp}{
  short=MDP,
  long=Markov decision process,
  short-plural-form=MDPs,
  long-plural-form=Markov decision processes
}
\DeclareAcronym{sg}{
  short=SG,
  long=stochastic game,
  short-plural-form=SGs,
  long-plural-form=stochastic games
}
\DeclareAcronym{ppg}{
  short=PPG,
  long=probabilistic parity game
}
\DeclareAcronym{imc}{
  short=IMC,
  long=interactive Markov chain,
}
\DeclareAcronym{ma}{
  short=MA,
  long=Markov automaton,
  short-plural-form=MA,
  long-plural-form=Markov automata
}
\DeclareAcronym{mra}{
  short=MRA,
  long=Markov reward automaton,
  short-plural-form=MRA,
  long-plural-form=Markov reward automata
}
\DeclareAcronym{pta}{
  short=PTA,
  long=probabilistic timed automaton,
  short-plural-form=PTA,
  long-plural-form=probabilistic timed automata
}
\DeclareAcronym{qvbs}{
  short=QVBS,
  long=Quantitative Verification Benchmark Set
}
\DeclareAcronym{sha}{
  short=SHA,
  long=stochastic automaton,
  short-plural-form=SHA,
  long-plural-form=stochastic timed automata
}
\DeclareAcronym{sta}{
  short=STA,
  long=stochastic timed automaton,
  short-plural-form=STA,
  long-plural-form=stochastic timed automata
}
\DeclareAcronym{gspn}{
  short=GSPN,
  long=generalized stochastic Petri net
}
\DeclareAcronym{dft}{
  short=DFT,
  long=dynamic fault tree
}
\DeclareAcronym{bn}{
  short=BN,
  long=Bayesian network
}
\DeclareAcronym{sma}{
  short=SMA,
  long=symbolic Markov automaton,
  short-plural-form=SMA,
  long-plural-form=symbolic Markov automata
}
\DeclareAcronym{spa}{
  short=SPA,
  long=symbolic probabilistic automaton,
  short-plural-form=SPA,
  long-plural-form=symbolic probabilistic automata
}
\DeclareAcronym{nsma}{
  short=NSMA,
  long=network of symbolic Markov automata,
  short-plural-form=NSMA,
  long-plural-form=networks of symbolic Markov automata
}
\DeclareAcronym{nspa}{
  short=NSPA,
  long=network of symbolic probabilistic automata,
  short-plural-form=NSPA,
  long-plural-form=networks of probabilistic automata
}
\DeclareAcronym{srm}{
  short=SRM,
  long=symbolic reward model,
}

\DeclareAcronym{bisim}{
  short=BM,
  long=bisimulation minimization
}
\DeclareAcronym{sat}{
  short=SAT,
  long=satisfiability
}
\DeclareAcronym{smt}{
  short=SMT,
  long=satisfiability modulo theories
}
\DeclareAcronym{lia}{
  short=LIA,
  long=linear integer arithmetic
}
\DeclareAcronym{lra}{
  short=LRA,
  long=linear real arithmetic
}
\DeclareAcronym{api}{
  short=API,
  long=application programming interface
}
\DeclareAcronym{lp}{
  short=LP,
  long=linear program
}
\DeclareAcronym{milp}{
  short=MILP,
  long=mixed-integer linear program
}

\DeclareAcronym{ltl}{
  short=LTL,
  long=linear temporal logic
}
\DeclareAcronym{pltl}{
  short=PLTL,
  long=probabilistic linear temporal logic
}
\DeclareAcronym{ctl}{
  short=CTL,
  long=computation tree logic
}
\DeclareAcronym{pctl}{
  short=PCTL,
  long=probabilistic computation tree logic
}
\DeclareAcronym{pctlstar}{
  short=PCTL$^{*}$,
  long=probabilistic computation tree logic
}
\DeclareAcronym{csl}{
  short=CSL,
  long=continuous stochastic logic
}

%% file: preamble/macros.tex
\newcommand{\eg}{e.\,g.}
\newcommand{\ie}{i.\,e.}

\newcommand{\nocheckmark}{\ensuremath{\times}}

\newcommand{\tuple}[1]{\ensuremath{\left\langle #1 \right\rangle}}

\newcommand{\tool}[1]{\textsc{#1}}

\newcommand{\prism}{\tool{Prism}}

\newcommand{\mrmc}{\tool{Mrmc}}

\newcommand{\storm}{\tool{Storm}}

\newcommand{\modesttoolset}{\tool{Modest Toolset}}
\newcommand{\modesttool}{\tool{Modest}}

\newcommand{\fig}{\tool{Fig}}

\newcommand{\epmc}{\tool{Epmc}}

\newcommand{\prophesy}{\tool{PROPhESY}}

\newcommand{\zthree}{\tool{Z3}}
\newcommand{\gurobi}{\tool{Gurobi}}

\newcommand{\glpk}{\tool{glpk}}

\newcommand{\cudd}{\tool{CUDD}}
\newcommand{\sylvan}{\tool{Sylvan}}

\newcommand{\smtrat}{\tool{Smt-Rat}}
\newcommand{\carl}{\tool{CArL}}
\newcommand{\mathsat}{\tool{MathSat}}
\newcommand{\gmp}{\tool{GMP}}

\newcommand{\gmmxx}{\tool{Gmm++}}
\newcommand{\eigen}{\tool{Eigen}}
\newcommand{\stormpy}{\tool{Stormpy}}

\newcommand{\mcsta}{\tool{mcsta}}

\newcommand{\engine}[1]{\textsf{#1}}

\newcommand{\lang}[1]{\textsc{#1}}

\newcommand{\prismlang}{\lang{Prism}}

\newcommand{\jani}{\lang{Jani}}

\newcommand{\galileo}{\lang{Galileo}}
\newcommand{\pgcl}{\lang{pGCL}}

\newcommand{\clanguage}{\lang{C}}

\newcommand{\cpp}{C++}
\newcommand{\python}{Python}

\newcommand{\pnml}{\lang{PNML}}

\newcommand{\smtlib}{\lang{SMT-LIB}}

\newcommand{\probname}[1]{\textsc{#1}}

\newcommand{\smt}{\probname{Smt}}

\newcommand{\milp}{\probname{Milp}}

\definecolor{color1}{RGB}{55,126,184} % lightblueish
\definecolor{color2}{RGB}{228,26,28} % red
\definecolor{color3}{RGB}{77,175,74} % green
\definecolor{color4}{RGB}{152,78,163} % purple
\definecolor{color5}{RGB}{255,127,0} % orange
\definecolor{color6}{rgb}{0.5, 1.0, 0.83} % aquamarine
\definecolor{color7}{rgb}{1.0, 0.0, 1.0} % magenta
\definecolor{color8}{rgb}{0.66, 0.66, 0.66} % gray

\newlength{\quantileplotwidth}
\newlength{\quantileplotheight}
\newcommand{\quantileplotlegendpos}{south east}
\newcommand{\quantileplotlegendcols}{1}
\tikzset{tool/.code={%
		\ifthenelse{\equal{#1}{mcsta}}{\tikzset{color1}}{}%, mark=x, mark options={thick}}}{}%
		\ifthenelse{\equal{#1}{Storm}}{\tikzset{color2}}{}%,mark=+, mark options={thick}}}{}%
		\ifthenelse{\equal{#1}{epmc}}{\tikzset{color5}}{}%, mark=*, mark size=1.5pt}}{}%
		\ifthenelse{\equal{#1}{PRISM}}{\tikzset{color3}}{}%, mark=asterisk}}{}%
		\ifthenelse{\equal{#1}{Storm.hybrid}}{\tikzset{color6}}{}%,mark=+, mark options={thick}}}{}%
		\ifthenelse{\equal{#1}{Storm.sparse}}{\tikzset{color2}}{}%, mark=x, mark options={thick}}}{}%
		\ifthenelse{\equal{#1}{Storm.sound}}{\tikzset{color3}}{}%, mark=*, mark size=1.5pt}}{}%
		\ifthenelse{\equal{#1}{Storm.dd}}{\tikzset{color4}}{}%, mark=*, mark size=1.5pt}}{}%
		\ifthenelse{\equal{#1}{Storm.exact}}{\tikzset{color5}}{}%, mark=*, mark size=1.5pt}}{}%
		\ifthenelse{\equal{#1}{Storm.dft}}{\tikzset{color6}}{}%, mark=*, mark size=1.5pt}}{}%
		\ifthenelse{\equal{#1}{Storm.ddbisim}}{\tikzset{color7}}{}%, mark=asterisk}}{}%
		\ifthenelse{\equal{#1}{Storm.automatic}}{\tikzset{color1}}{}%, mark=asterisk}}{}%
		\ifthenelse{\equal{#1}{Storm.best}}{\tikzset{color8}}{}%, mark=asterisk}}{}%
}}
\newcommand{\quantileplot}[4]{%
	\begin{tikzpicture}
	\begin{axis}[
	width=\quantileplotwidth,
	height=\quantileplotheight,
	xmin=1,
	ymax=2300,
	ymin=1,
	ymode=log,
	axis x line=bottom,
	axis y line=left,
	ytick= {1, 6, 60, 600, 1200, 1800 },
	yticklabels={$\le$1, 6, 60, 600, 1200, 1800},
	xlabel=solved instances (out of #4),
	ylabel=seconds (slowest instance),
	x label style={at={(axis description cs:0.5,-0.05)},anchor=north},
	yticklabel style={font=\scriptsize},
	xticklabel style={font=\scriptsize},
	legend pos=\quantileplotlegendpos,
	legend columns=\quantileplotlegendcols,
	legend style={nodes={scale=0.75, transform shape},inner sep=1pt},
	every axis plot/.append style={ultra thick},
	legend cell align={left}
	]
	\foreach \tool in {#2}{%
		\edef\loopbody{
			\noexpand\addplot[tool=\tool] table [x=n,y=\tool shifted, col sep=semicolon] {#1};
		}
		\loopbody
	}
	\draw[densely dotted] (axis cs: 0,6) -- (axis cs: 100,6);
	\draw[densely dotted] (axis cs: 0,60) -- (axis cs: 100,60);
	\draw[densely dotted] (axis cs: 0,600) -- (axis cs: 100,600);
	\draw[densely dotted] (axis cs: 0,1200) -- (axis cs: 100,1200);
	\draw[densely dotted] (axis cs: 0,1800) -- (axis cs: 100,1800);
	\legend{#3}
	\end{axis}
	\end{tikzpicture}%
}

\newlength{\scatterplotsize}
\newcommand{\scatterplot}[4]{%
	\begin{tikzpicture}
	\begin{axis}[
	width=\scatterplotsize,
	height=\scatterplotsize,
	axis equal image,
	xmin=0.04,
	ymin=0.04,
	ymax=6000,
	xmax=25000,
	xmode=log,
	ymode=log,
	axis x line=bottom,
	axis y line=left,
	xtick={1,6,60,600,1800},
	xticklabels={1,6,60,600,1800},
	extra x ticks = {4000,8000,16000},
	extra x tick labels = {TO,ERR,INC},
	extra x tick style = {grid = major},
	ytick={1,6,60,600,1800},
	yticklabels={1,6,60,600,1800},
	extra y ticks = {4000},
	extra y tick labels = {n/a},
	extra y tick style = {grid = major},
	xlabel={\strut#3},
	xlabel style={yshift=5pt},
	ylabel=best of other tools,
	ylabel style={yshift=-0.4cm},
	yticklabel style={font=\scriptsize},
	xticklabel style={rotate=290,anchor=west,font=\scriptsize},
	legend pos=north west,
	legend columns=-1,
	legend style={nodes={scale=0.75, transform shape},inner sep=1.5pt,yshift=0.67cm,xshift=0.8cm},
	]
	
	\addplot[
	scatter,
	only marks,
	scatter/classes={
		dtmc={mark=square*,color1,mark size=1.25},
		ctmc={mark=diamond*,color3,mark size=1.75},
		mdp={mark=triangle*,color2,mark size=1.75},
		ma={mark=pentagon*,color4,mark size=1.50},
		pta={mark=*,color5,mark size=1.33}
	},
	scatter src=explicit symbolic
	]%
	table [col sep=semicolon,x=#2,y=other,meta=Type] {#1};
	\ifthenelse{\NOT\equal{#4}{false}}{\legend{DTMC, CTMC, MDP, MA, PTA}}{}
	\addplot[no marks] coordinates {(0.01,0.01) (3600,3600) };
	\addplot[no marks, densely dotted] coordinates {(0.01,0.1) (360,3600)};
	\end{axis}
	\end{tikzpicture}
}
\newcommand{\scatterplotstorm}[6]{%
	\begin{tikzpicture}
	\begin{axis}[
	width=\scatterplotsize,
	height=\scatterplotsize,
	axis equal image,
	xmin=0.04,
	ymin=0.04,
	ymax=18000,
	xmax=18000,
	xmode=log,
	ymode=log,
	axis x line=bottom,
	axis y line=left,
	xtick={1,6,60,600,1800},
	xticklabels={1,6,60,600,1800},
	extra x ticks = {4000,7000,12000},
	extra x tick labels = {OOR,NS,INC},
	extra x tick style = {grid = major},
	ytick={1,6,60,600,1800},
	yticklabels={1,6,60,600,1800},
	extra y ticks = {4000,7000,12000},
	extra y tick labels = {OOR,NS,INC},
	extra y tick style = {grid = major},
	xlabel={#3},
	xlabel style={yshift=16pt},
	ylabel={#5},
	ylabel style={yshift=-0.4cm},
	yticklabel style={font=\scriptsize},
	xticklabel style={rotate=290,anchor=west,font=\scriptsize},
	legend pos=north west,
	legend columns=-1,
	legend style={nodes={scale=0.75, transform shape},inner sep=1.5pt,yshift=0.67cm,xshift=0.8cm},
	]
	
	\addplot[
	scatter,
	only marks,
	scatter/classes={
		dtmc={mark=square*,color1,mark size=1.25},
		ctmc={mark=diamond*,color3,mark size=1.75},
		mdp={mark=triangle*,color2,mark size=1.75},
		ma={mark=pentagon*,color4,mark size=1.50},
		pta={mark=*,color5,mark size=1.33}
	},
	scatter src=explicit symbolic
	]%
	table [col sep=semicolon,x=#2,y=#4,meta=Type] {#1};
	\ifthenelse{\NOT\equal{#6}{false}}{\legend{DTMC, CTMC, MDP, MA, PTA}}{}
	\addplot[no marks] coordinates {(0.01,0.01) (3600,3600) };
	\addplot[no marks, densely dotted] coordinates {(0.01,0.1) (360,3600)};
	\addplot[no marks, densely dotted] coordinates {(0.1,0.01) (3600,360)};
	\end{axis}
	\end{tikzpicture}
}

\newcommand{\scatterplotstormfastesty}[6]{%
	\begin{tikzpicture}
	\begin{axis}[
	width=\scatterplotsize,
	height=\scatterplotsize,
	axis equal image,
	xmin=0.04,
	ymin=0.04,
	ymax=18000,
	xmax=18000,
	xmode=log,
	ymode=log,
	axis x line=bottom,
	axis y line=left,
	xtick={1,6,60,600,1800},
	xticklabels={1,6,60,600,1800},
	extra x ticks = {4000,7000,12000},
	extra x tick labels = {OOR,NS,INC},
	extra x tick style = {grid = major},
	ytick={1,6,60,600,1800},
	yticklabels={1,6,60,600,1800},
	extra y ticks = {4000},
	extra y tick labels = {not solved},
	extra y tick style = {grid = major},
	xlabel={#3},
	xlabel style={yshift=16pt},
	ylabel={#5},
	ylabel style={yshift=-0.4cm},
	yticklabel style={font=\scriptsize},
	xticklabel style={rotate=290,anchor=west,font=\scriptsize},
	legend pos=north west,
	legend columns=-1,
	legend style={nodes={scale=0.75, transform shape},inner sep=1.5pt,yshift=0.67cm,xshift=0.8cm},
	]
	
	\addplot[
	scatter,
	only marks,
	scatter/classes={
		dtmc={mark=square*,color1,mark size=1.25},
		ctmc={mark=diamond*,color3,mark size=1.75},
		mdp={mark=triangle*,color2,mark size=1.75},
		ma={mark=pentagon*,color4,mark size=1.50},
		pta={mark=*,color5,mark size=1.33}
	},
	scatter src=explicit symbolic
	]%
	table [col sep=semicolon,x=#2,y=#4,meta=Type] {#1};
	\ifthenelse{\NOT\equal{#6}{false}}{\legend{DTMC, CTMC, MDP, MA, PTA}}{}
	\addplot[no marks] coordinates {(0.01,0.01) (3600,3600) };
	\addplot[no marks, densely dotted] coordinates {(0.01,0.1) (360,3600)};
	\addplot[no marks, densely dotted] coordinates {(0.1,0.01) (3600,360)};
	\end{axis}
	\end{tikzpicture}
}

\newcommand*{\FormatDigit}[1]{\textcolor{blue}{#1}}
\lstdefinestyle{FormattedNumber}{%
    literate={0}{{\FormatDigit{0}}}{1}%
             {1}{{\FormatDigit{1}}}{1}%
             {2}{{\FormatDigit{2}}}{1}%
             {3}{{\FormatDigit{3}}}{1}%
             {4}{{\FormatDigit{4}}}{1}%
             {5}{{\FormatDigit{5}}}{1}%
             {6}{{\FormatDigit{6}}}{1}%
             {7}{{\FormatDigit{7}}}{1}%
             {8}{{\FormatDigit{8}}}{1}%
             {9}{{\FormatDigit{9}}}{1}%
             {.0}{{\FormatDigit{.0}}}{2}% Following is to ensure that only periods
             {.1}{{\FormatDigit{.1}}}{2}% followed by a digit are changed.
             {.2}{{\FormatDigit{.2}}}{2}%
             {.3}{{\FormatDigit{.3}}}{2}%
             {.4}{{\FormatDigit{.4}}}{2}%
             {.5}{{\FormatDigit{.5}}}{2}%
             {.6}{{\FormatDigit{.6}}}{2}%
             {.7}{{\FormatDigit{.7}}}{2}%
             {.8}{{\FormatDigit{.8}}}{2}%
             {.9}{{\FormatDigit{.9}}}{2}%
             {\ }{{ }}{1}% handle the space
             ,
   basicstyle=\ttfamily,%  Optional to use this
}

%% file: contents/title.tex
\title{The Probabilistic Model Checker \storm}

\author{Christian Hensel \and Sebastian Junges \and Joost-Pieter Katoen \\ \and Tim Quatmann \and Matthias Volk}
\authorrunning{C. Hensel et al.}
\institute{RWTH Aachen University, Aachen, Germany}

\institute{
          C. Hensel \at
          RWTH Aachen University, Aachen, Germany\\
              \email{dehnert@cs.rwth-aachen.de}
          \and
           S. Junges \at
           University of California, Berkeley, California, USA \\
              \email{sjunges@berkeley.edu}
          \and
           J.-P. Katoen \at
            RWTH Aachen University, Aachen, Germany\\
              \email{katoen@cs.rwth-aachen.de}
          \and
           T. Quatmann \at
            RWTH Aachen University, Aachen, Germany\\
              \email{tim.quatmann@cs.rwth-aachen.de}
          \and
           M. Volk \at
            RWTH Aachen University, Aachen, Germany\\
              \email{matthias.volk@cs.rwth-aachen.de}
}

\iftoggle{TR}{\date{}}{\date{Received: date / Accepted: date}}
\maketitle

%% file: contents/abstract.tex
%%\begin{abstract}
%%We present the state-of-the-art probabilistic model checker \storm{}.
%%\storm{} features the analysis of discrete- and continuous-time variants of both Markov chains and MDPs, which may be described using, e.g., the JANI modelling language, dynamic fault trees or generalised stochastic Petri nets. \todo{prism language?}
%%Its modular set-up in which solvers and symbolic engines can easily be exchanged allows to adapt research successes.\todo{word missing?}
%%Its Python API allows for rapid prototyping by encapsulating \storm's fast and scalable algorithms. 
%%\end{abstract}

\begin{abstract}
We present the probabilistic model checker \storm{}.
\storm{} supports the analysis of discrete- and continuous-time variants of both Markov chains and Markov decision processes.
\storm{} has three major distinguishing features.
It supports multiple input languages for Markov models, including the \jani{} and \prism{} modeling languages, dynamic fault trees, generalized stochastic Petri nets and the probabilistic guarded command language.
It has a modular set-up in which solvers and symbolic engines can easily be exchanged.
Its Python API allows for rapid prototyping by encapsulating \storm's fast and scalable algorithms. 
This paper reports on the main features of \storm{} and explains how to effectively use them.
A description is provided of the main distinguishing functionalities of \storm{}.
Finally, an empirical evaluation of different configurations of \storm{} on the QComp 2019 benchmark set is presented.
\end{abstract}

%% file: contents/introduction.tex
\section{Introduction}
The verification of systems involving stochastic uncertainty is a prominent research challenge.
Among the many techniques is probabilistic model checking, a mature technique that grew out of model checking.
\begin{figure}[t]
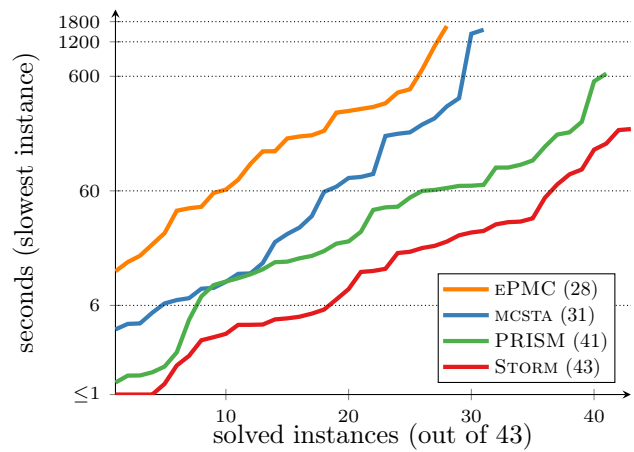

	\begin{center}
	\quantileplot{contents/plotdata/qcomp/specificgeneralpurpose.csv}{epmc,mcsta,PRISM,Storm}{\tool{ePMC} (28),\tool{mcsta} (31),\tool{PRISM} (41),\tool{Storm} (43)}{43}
	\end{center}
	\caption{Runtime comparison of general-purpose probabilistic model checkers taken from the QComp 2019 report~\cite{DBLP:conf/tacas/HahnHHKKKPQRS19}$^{*}$}
	\footnotesize{$^{*}$: Licensed under~Creative Commons Attribution 4.0 International License: \url{http://creativecommons.org/licenses/by/4.0/}.}
	% Regarding Licensing:
	% [The license] permits use, sharing, adaptation, distribution and reproduction in any medium or format, as long as you give appropriate credit to the original author(s) and the source, *provide a link to the Creative Commons license* and indicate if changes were made.
	\label{fig:QuantilePlotsTweaked}
\end{figure}

A model checker takes the formal system model and the formal property as inputs and, somewhat simplifying, returns one of three results, see \cref{fig:introduction_overview_model_checking}.
It reports that the property holds or is violated, and these reports are---given a correct implementation---guaranteed to be correct.
%In the latter case, counterexamples can be synthesized that explain the cause of the violation as succinctly as possible, an effective means to convince engineers of a design problem.
The third outcome is that the model checker ran out of computational resources.
Model checking has written numerous success stories~\cite{DBLP:journals/cacm/BallLR11,DBLP:journals/cacm/Holzmann14}, and major contributors Edmund M.\ Clarke, E.\ Allen Emerson and Joseph Sifakis were awarded the Turing Award in 2007.
\emph{Probabilistic model checking} extends traditional model checking with tools and techniques for the analysis of systems involving random phenomena or other forms of behavior that can be approximated by randomization.
Alur, Henzinger and Vardi~\cite{DBLP:journals/siglog/AlurHV15} state: ``A promising new direction in formal methods is probabilistic model checking, with associated tools for quantitative evaluation of system performance along with correctness''.
Distributed algorithms and communication protocols are natural examples, as they often use randomization to efficiently break symmetry.
Another example are cyber-physical systems that tightly integrate software and hardware such as sensors, actors and micro-controllers. 
In particular, sensor readings may be noisy, actors may not always have the same effects, and physical components may fail.
Other domains that give rise to models involving probabilistic aspects include, \eg{} security protocols and systems biology.
All these systems are naturally mapped to Markov models, and probabilistic model checking takes exactly such models as input.

Probabilistic model checking is not new.
Initial theoretical results and algorithms for Markov chains~\cite{DBLP:conf/rtss/HanssonJ89,DBLP:journals/fac/HanssonJ94} and Markov decision processes~\cite{DBLP:conf/focs/Vardi85,DBLP:conf/focs/CourcoubetisY88} were provided about thirty years ago.
%The use of symbolic data structures led to the first serious tool support \cite{DBLP:conf/icalp/BaierCHKR97,DBLP:conf/arts/Hartonas-GarmhausenCC99}.
First tool support using explicit~\cite{Fredlund1994TheTA} and symbolic data structures~\cite{DBLP:conf/icalp/BaierCHKR97,DBLP:conf/arts/Hartonas-GarmhausenCC99} followed.
Tool realizations for continuous-time Markov chains appeared shortly thereafter \cite{DBLP:conf/tacas/HermannsKMS00}.
\prism{} evolved as one of the main probabilistic model checkers\footnote{Resulting in the Haifa Verification Conference 2016 Award.} covering all these models in a symbolic way~\cite{DBLP:conf/tacas/KwiatkowskaNP02}. 
In more recent years, tool support extended to cover probabilistic real-time and hybrid systems, as well as multi-player games. 

Meanwhile, research in probabilistic model checking continued, changed directions, and progressed in new application areas. 
The diversity of this field motivated the development of a modular and adaptive model checker, called \storm{}.
\storm{}'s main aim is to be a performant, easy-extendible platform supplying various probabilistic model checking algorithms.
After five years of development, \storm{} was released as open-source project in 2017~\cite{DBLP:conf/cav/DehnertJK017}.
Despite its relative young age, \storm{} has established the following in pursuit of its original goals:
%% \storm{} has shown to be successful in pursue of its original goals:
\begin{itemize}
\item 
In the first edition of QComp~\cite{DBLP:conf/tacas/HahnHHKKKPQRS19}, \storm{} compared favorably with other model checkers. Consider the \emph{quantile-plot} in \cref{fig:QuantilePlotsTweaked}. 
The quantile plot expresses how many benchmark instances (on the x-axis) \emph{each} were solved in at most the time given on the y-axis.
In other words, the point $\tuple{x, y}$ is contained in the quantile plot for tool \engine{c} if the \emph{maximal} runtime when using \engine{c} on the $x$ fastest solved instances (for \engine{c}) is $y$ seconds.
\storm{} solved more instances, and was generally faster in solving these instances. We elaborate these results in \cref{sec:qcomp}.
\item 
\storm's modularity paid off in various occasions: The tool has been adapted to include various novel variants to the typical value iteration algorithm, and has been extended with parameter synthesis for probabilistic systems and multi-objective model checking. In many of these areas, \storm{} has helped to push the state-of-the-art considerably. We elaborate these results in \cref{sec:features}.
\end{itemize}

In this paper, we report on \storm's main features and how to use them.
We start with a very quick overview introducing \storm{} before elaborating the supported models and properties.
We survey \storm's most prominent building blocks and unique features in greater detail, and discuss the possibilities to interface with these features in \storm.
Finally, we report on its internal tool architecture, and provide some empirical evaluation of the main configurations of \storm{} on the QComp 2019 benchmark set.

A video tutorial covering \storm{} and some of its core features is available at
\begin{center}
\url{http://stormchecker.org/video-tutorial}.
\end{center}
\section{Storm in a Nutshell}
Research to advance concepts and methods for probabilistic model checking often combines key routines and a variety of essential model checking algorithms. \storm{} delivers these. Some main characteristic features of \storm{} that help to push the state-of-the-art in probabilistic model checking are that \storm 
\begin{itemize}
\item 
contains efficient implementations of well-known and mature model checking algorithms for discrete-time and continuous-time Markov chains and Markov decision processes, but also for the more general \emph{Markov automata}~\cite{DBLP:conf/lics/EisentrautHZ10}, a model containing probabilistic branching, non-determinism, and exponentially distributed delays\footnote{Markov automata can be used to provide a compositional semantics to modeling formalisms such as arbitrary generalized stochastic Petri nets~\cite{DBLP:conf/apn/EisentrautHK013}, dynamic fault trees~\cite{DBLP:conf/atva/BoudaliCS07}, and AADL extended with the error annex~\cite{DBLP:journals/cj/BozzanoCKNNR11}};
\item
supports \emph{explicit-state} and \emph{symbolic} (BDD-based) model checking as well as a \emph{mixture} of these modes to handle a wider range of models;
\item 
has a \emph{modular} set-up, enabling the easy exchange of different solvers and distinct decision diagram packages; its current release supports about 15 solvers, and two BDD packages. %{\tt CUDD}~\cite{cudd_website} and multi-threaded \sylvan~\cite{DBLP:phd/basesearch/Dijk16};
\item 
extends probabilistic model checking with the possibility of generating (high-level) counterexample~\cite{DBLP:conf/atva/DehnertJWAK14}, synthesizing permissive schedulers~\cite{DBLP:journals/corr/DragerFK0U15}, symbolic bisimulation minimization~\cite{DBLP:journals/sttt/DijkP18,DBLP:conf/gi/Wimmer11} as well as game-based abstraction of infinite-state MDPs~\cite{DBLP:phd/de/Wachter2011}.
\item
offers the possibility to improve the reliability of model checking by supporting exact rational arithmetic using recent techniques~\cite{DBLP:conf/fmcad/BauerMCS017}, and techniques to avoid premature termination of value iteration~\cite{DBLP:conf/cav/QuatmannK18}.
\item 
supports advanced properties such as multi-objective model checking~\cite{DBLP:conf/tacas/ForejtKNPQ11,DBLP:conf/atva/ForejtKP12,DBLP:conf/cav/QuatmannJK17}, efficient algorithms for conditional probabilities and rewards~\cite{DBLP:conf/tacas/BaierKKM14}, and long-run averages on MDPs~\cite{DBLP:conf/lics/Alfaro98,DBLP:conf/cav/AshokCDKM17} and MAs~\cite{DBLP:conf/tacas/ButkovaWH17}.
\storm{} also contains (the essential building blocks) for handling parametric models such as~\cite{DBLP:conf/ictac/Daws04,DBLP:conf/atva/QuatmannD0JK16,DBLP:conf/atva/SpelJK19};
\end{itemize}
\storm{} can also be used to investigate the application of model checking in novel domains: In particular,
\begin{itemize}
\item
\storm{} supports \emph{various native input formats}: the \prismlang{} and \jani{} languages, generalized stochastic Petri nets, dynamic fault trees, and conditioned probabilistic programs. 
This support makes it easier to apply probabilistic model checking, and amounts not to just providing another parser; state-space reduction and generation techniques as well as analysis algorithms are partly tailored to these modeling formalisms;
\item
besides a command line interface with many optional arguments, \storm{} provides a \emph{Python API} facilitating easy and rapid prototyping of other tools using the engines and algorithms of~\storm;
\item
it provides advanced approaches to model checking (see above) and good performance in terms of verification speed and memory footprint, cf.\ \cref{fig:QuantilePlotsTweaked}, under one roof.
\end{itemize}

\paragraph{How does \storm{} relate to other probabilistic model checkers?}
\storm{} has not reinvented the wheel, but has rather been inspired and learned from the successes of in particular \prism{}~\cite{DBLP:conf/cav/KwiatkowskaNP11} and the explicit model checker \mrmc{}~\cite{DBLP:journals/pe/KatoenZHHJ11}.
Like its main competitors \prism, \mcsta~\cite{DBLP:conf/tacas/HartmannsH14}, and \epmc~\cite{DBLP:conf/fm/HahnLSTZ14}, \storm{} relies on numerical and symbolic computations.
Although many functionalities are covered by \storm, there are some significant areas that \storm{} has not been extended to. 
It does not support discrete-event simulation against temporal logic formulas, known as statistical model checking~\cite{DBLP:conf/isola/LarsenL16,DBLP:journals/tomacs/AghaP18}.
\storm{} does not support LTL model checking (as supported by \epmc{} and \prism), does not support probabilistic timed automata (as supported by \mcsta{} and \prism), has no equivalent of \prism's \engine{hybrid} engine (a crossover between full MTBDD and \storm's \engine{hybrid} engine), and does not support the analysis of stochastic games.
A longer survey of both features and performance of the various model checkers can be found in \cite{DBLP:conf/tacas/HahnHHKKKPQRS19,qcomp2020}.
A detailed comparison between \storm{}, \epmc{}, \mcsta{}, and \prism{} is given in~\cite{DBLP:phd/dnb/Hensel18}.

%% file: contents/probmodelcheck.tex
\section{Probabilistic Model Checking with \storm{}}
\label{sec:probmc}
\begin{figure}
  \begin{center}
\scalebox{0.9}{
	\begin{tikzpicture}
        \tikzset{input/.style={draw, ellipse, minimum width=2.5cm, minimum height=.7cm}}
        \tikzset{output/.style={input}}
        \tikzset{astep/.style={draw, rectangle, rounded corners, minimum width=2.5cm, minimum height=.7cm}}
        \tikzset{flow_edge/.style={->,thick}}

        \renewcommand{\tikzhorizdist}{1.4cm}

        \node[astep, align=center] (modelchecking) {model checking\\{\footnotesize\cref{sec:methods}}};
	
        \node[input, above left=\tikzhorizdist and 0.4cm of modelchecking, align=center] (model) {model\\\footnotesize{\cref{sec:modeltypes}}};
        \node[input, above=\tikzhorizdist of model, align=center] (system) {system description\\
        \footnotesize{\cref{sec:languages}}};
        \node[input, above=0.5*\tikzhorizdist of system] (realsystem) {system};
        \node[input, above right=\tikzhorizdist and 0.4cm of modelchecking, align=center] (formulae) {properties\\{\footnotesize\cref{sec:properties}}};
        \node[input, above=\tikzhorizdist of formulae] (requirements) {requirements};
        
        \node[output, below left=1.1*\tikzhorizdist and 0.6cm of modelchecking] (satisfied) {satisfied};
        \node[output, below right=1.1*\tikzhorizdist and 0.6cm of modelchecking, label=below:{+ counterexample}] (violated) {violated};
        \node[output, dashed, below=0.6*\tikzhorizdist of modelchecking, align=center] (oor) {out of\\resources};

        \draw[flow_edge] (system) edge node[auto] {translates to} (model);
        \draw[flow_edge] (model) edge (modelchecking);
        \draw[flow_edge] (requirements) edge node[auto,swap] {formalizing} (formulae);
        \draw[flow_edge] (formulae) edge (modelchecking);
        \draw[flow_edge] (modelchecking) edge (satisfied);
        \draw[flow_edge, dashed] (modelchecking) edge (oor);
        \draw[flow_edge] (modelchecking) edge (violated);
        \draw[flow_edge] (realsystem) edge node[auto] {modeling} (system);
	\end{tikzpicture}  }
  \end{center}
  \caption{Overview of the model checking approach~\cite{DBLP:books/daglib/0020348}.}
  \label{fig:introduction_overview_model_checking}
\end{figure}
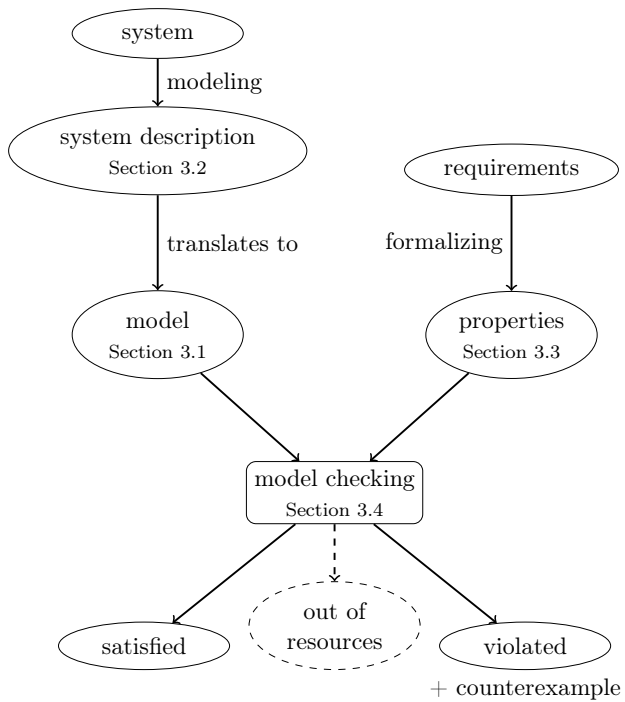

%In the context of probabilistic model checking, the state space explosion problem becomes even more pressing.
%Current techniques mostly require an in-memory representation of the full system under consideration.
%At the same time, the numerical solution techniques that are at the heart of probabilistic model checking become substantially more computationally expensive as the model grows.
%In 2008, in a retrospective on 25 years of model checking~\cite{DBLP:conf/spin/Clarke08}, Edmund Clarke states that probabilistic model checking is an important challenge for the future that ``require[s] major breakthroughs in order to become sufficiently practical for widespread use in industry.''

We give a gentle introduction to probabilistic model checking\footnote{Readers familiar with probabilistic model checking may safely skip this section.} with \storm, clarifying the different parts as outlined in \cref{fig:introduction_overview_model_checking}. For surveys and more formal introductions to probabilistic model checking, we refer to~\cite{DBLP:books/daglib/0020348,DBLP:conf/lics/Katoen16,DBLP:reference/mc/BaierAFK18}.

\subsection{Model Types}
\label{sec:modeltypes}
\storm{} supports the analysis of several different formalisms. They differ regarding
\begin{enumerate*}[label=(\roman*)]
  \item their notion of time and
  \item whether or not nondeterministic choices are allowed.
\end{enumerate*}
\Cref{tab:model_types} shows a categorization of the models supported by \storm{} along the two dimensions.
In a third dimension, \storm{} supports partially observable models, in which the way nondeterminism is resolved is restricted.

Discrete-time models abstract from timing behavior by viewing the progression of time in terms of discrete steps.
In contrast, continuous-time models use real numbers to model the flow of time and therefore have a dense notion of time.
Deterministic models (also referred to as Markov chains from now on) behave purely probabilistically.
Dually, in MDPs and MAs, nondeterministic choices can be used to model, for instance, the interaction with an adversarial environment or underspecification of the model with the goal to synthesize the optimal concrete system.
In general, all model types can be enriched with cost structures.
Together with the probabilities in the model this allows for reasoning over, for instance, expected costs until a certain goal is reached.
Rather than providing formal definitions, we will illustrate a typical use-case for each model type.
\begin{table}
  \begin{center}
    \begin{tabular}{r|cc}
%      \toprule
      & deterministic & nondeterministic \\
%      \cmidrule{2-3}
      \midrule
      discrete time & DTMCs & MDPs and PAs \\
%      \rowcolor{black-10}
      continuous time & CTMCs & MAs \\
%      \bottomrule
    \end{tabular}
  \end{center}
  \caption{Overview of model types.}
  \label{tab:model_types}
\end{table}

\subsubsection{Discrete-time Markov Chains (DTMCs)}
We start with the simplest model.
In DTMCs~\cite{DBLP:books/daglib/0095301} every state is equipped with a single probability distribution over successor states.
The evolution of the system therefore is fully probabilistic in the sense that it is governed only by repeated randomized trials.
A famous example that can be captured in terms of a DTMC is the Herman protocol~\cite{DBLP:journals/fac/KwiatkowskaNP12}.
The general setting is this: a ring consisting of identical processes that each start either with a token or without one.
If more than one process holds a token the protocol is in an \emph{unstable} state.
The goal is to reach a configuration in which \emph{exactly one} token remains, a situation called a stable configuration.
This problem cannot be solved by deterministic algorithms and randomization is crucial.
Herman's protocol uses synchronous, unidirectional communication and can be shown to eventually reach a stable configuration with probability~1.

\subsubsection{Continuous-time Markov Chains (CTMCs)}
CTMCs~\cite{DBLP:books/daglib/0095301} extend DTMCs with a continuous notion of time.
Here, the sojourn time of the system in a state is also determined by a random experiment.
More specifically, the time is sampled according to a negative exponential distribution.
The transitions between states happen just like for DTMCs, \ie, governed by the associated probability distributions.
Examples for CTMCs can be found in, for instance, systems biology~\cite{DBLP:journals/tcsb/CalderVGO06}.
In this work, they are used to analyze the effect of concentrations of proteins and reaction rates on signal transduction pathways.
In other words, the model combines discrete aspects (the molecule concentration) and continuous aspects (time).
Here, not only the probabilistic but also the timing effects are important: since both the underlying chemical reactions as well as the spatial distribution of molecules take time, fundamental questions like ``what is the probability that the concentration of $X$ is high after $10$ seconds?'' require a proper modeling of~time.

\subsubsection{Markov Decision Processes (MDPs)}
MDPs~\cite{Put94} extend DTMCs with nondeterminic actions.
That is, instead of a single distribution governing the successor states, the system can nondeterministically select between several actions, each identifying a different distribution.
After a selection has been made, the successor states are resolved probabilistically and in the successor state a new selection process is initiated.
As already mentioned, nondeterminism can be used to model the possible interaction with an adversarial environment.
An important example for this are distributed protocols.
Such protocols are often randomized to efficiently break symmetry.
However, because of their distributed nature, the progress of the processes is not synchronized and they may be scheduled differently.
A well-known example is the randomized consensus algorithm by Aspnes and Herlihy~\cite{DBLP:conf/cav/KwiatkowskaNS01}.
In this protocol, the participating processes repeatedly modify a shared global counter based on the outcome of a coin flip until the whole system agrees on one of two outcomes, \ie,~consensus has been reached.
To faithfully model the protocol, nondeterminism can be used to account for the missing information about the scheduling of the competing accesses to the counter.
Probabilistic automata (PAs)~\cite{DBLP:journals/njc/SegalaL95} extend MDPs with a more flexible action labeling. % with action labels, that allow for slightly more flexible modeling. 

\subsubsection{Markov Automata (MAs)}
Finally, MAs~\cite{DBLP:conf/lics/EisentrautHZ10} extend PAs using the notion of continuous time that CTMCs use.
In probabilistic states no time passes, and the system nondeterministically selects one of the available probability distributions.
In Markovian states, an amount of time passes that is distributed in a negative exponential manner, as in CTMCs.
A well-known example is the stochastic job scheduling problem \cite{DBLP:conf/cav/QuatmannJK17}.
Here, the task is to schedule $n$ jobs with (different) exponential service times onto $k$ processors.
The processors are assumed to run a pre-emptive scheduling strategy: upon completion of any job, all $k$ processors can take over any of the remaining jobs.
The corresponding MA uses nondeterministic choices to model the assignment of jobs to processors whenever such a choice can be made.
Thus, the nondeterminism is used to \emph{underspecify} the concrete behavior.
Determining the job assignment that maximizes the probability for completion within a given time limit can thus be seen as synthesizing a scheduling policy that one would like to impose in the actual system.

\subsubsection{Partially observable MDPs (POMDPs)} 
Partially observable MDPs~\cite{Astrom65,DBLP:journals/ai/KaelblingLC98} are a popular extension that cater for a common issue with the analysis of MDPs. 
That analysis typically assumes that the nondeterminism can be resolved arbitrarily. The policy resolving the nondeterminism might, for example, depend on the internal state of a remotely running process.
Consequently, the policies that are synthesized by such an analysis are \emph{unrealistic}, and the verification results are too pessimistic.
Consider a game like mastermind, where the adversary has a trivial strategy if it knows the secret they have to guess. 
Intuitively, to analyze an adversary that has to find a secret, we must assume it cannot observe this secret. 
For a range of privacy, security, and robotic domains, we may instead assume that the adversary must decide based on system observations. In widespread examples~\cite{DBLP:journals/ai/KaelblingLC98}, the position of a robot is unknown and can only be determined by landmarks (such as doors), or the position of other agents in the same environment can only be observed if these agents are sufficiently close.

Formally, POMDPs extend MDPs by a set of observations, and label every state with a one of these observations. Extensions in which actions are labeled, or where states are labeled with distributions over observations can be reduced to this simpler case.

\subsection{Modeling Languages}
\label{sec:languages}
Markov models for practical purposes are often too large to denote explicitly, but may be described by various more powerful and concise  modeling languages.
Depending on the domain, different modeling languages are more or less suitable. 
Furthermore, the structure of the model is often more apparent from a symbolic description than on the state level. 
%For instance, the most well-known probabilistic model checker \prism~\cite{DBLP:conf/cav/KwiatkowskaNP11} supports the \prismlang{} language as well as an input format that explicitly enumerates the behavior of the system.
%These restrictions may make it necessary to transcode existing models into a different language, a process that is time-consuming and error-prone.
%Additionally, the tool that best fits the analysis might only offer modeling languages that do not fit the system well.
\storm{} therefore tries to support a variety of different input languages.
In order to be compatible with the wide-spread usage of \prism, the \prismlang{} language is supported.
For testing small models explicit enumeration of states and transitions is supported in two different formats.
Furthermore, \storm{} accepts models given in \jani~\cite{DBLP:conf/tacas/BuddeDHHJT17}, a modeling language that was devised in a joint effort across multiple tools (involving \epmc, \modesttool, \fig) in an attempt to unify the cluttered language landscape.
\storm{} supports three other modeling languages.
First, the user can input  generalized stochastic Petri nets (GSPNs)~\cite{DBLP:journals/tocs/MarsanCB84} specified in an extension of the Petri-net Markup Language \pnml, which is then translated to \jani{} automatically.
GSPNs are an important modeling formalism in dependability and performance evaluation.
Secondly, Dynamic Fault Trees (DFTs) are a means to specify the fault behavior of systems and is a reliability engineering formalism that is widely used in industry~\cite{DBLP:journals/csr/RuijtersS15}.
DFTs can be specified in the \galileo{} format~\cite{DBLP:conf/ftcs/SullivanDC99}.
Finally, a recent trend in the analysis of probabilistic systems is probabilistic programming~\cite{DBLP:conf/icse/GordonHNR14}. 
The latter refers to programs written in a probabilistic extension of regular programs.
An extension to imperative while programs is  \pgcl~\cite{DBLP:journals/scp/JifengSM97}, and can additionally be extended with statements expressing conditional reasoning~\cite{DBLP:journals/toplas/OlmedoGJKKM18}, an ingredient that is essential to describe inference as in Bayesian networks.
\storm{} can parse and translate programs written in \pgcl{} to \jani{}, which makes such programs amenable to existing probabilistic model checking techniques.

\subsection{Properties}
\label{sec:properties}
\storm{} offers support for a multitude of properties.
The most fundamental properties are reachability properties.
Intuitively, they ask for the probability with which a system reaches a certain state.
One may, e.g., ask 
\begin{itemize}
\item 
``is the probability to reach an unsafe state of the system less than $0.1$?''
\item 
``is the probability to reach a target within 20 steps at least $0.9$?''
\end{itemize}
For models involving nondeterministic choices, such an analysis will reason about all possible resolutions of nondeterminism and assert that the desired property holds \emph{in all cases}. Alternatively, an easy extension is to ask for \emph{some} resolution of the nondeterminism such that the property holds.
Besides asking for whether the probability meets some threshold, one may also ask ``what is the probability to reach an unsafe state of the system?''. 

As models can be equipped with cost structures, properties allow for retrieving, \eg
\begin{itemize}
  \item ``what is the expected number of coin flips until consensus has been reached?''
  \item ``what is the expected energy consumption after $t$ time units?''
  \item ``what is the expected molecule concentration at time point $t$?''
\end{itemize}
Further properties include temporal logic formulas based on PCTL~\cite{DBLP:journals/fac/HanssonJ94} and CSL~\cite{DBLP:journals/tocl/AzizSSB00,DBLP:journals/tse/BaierHHK03}, conditional probability and cost queries~\cite{DBLP:conf/tacas/BaierKKM14,DBLP:conf/tacas/Baier0KW17}, long-run average values~\cite{DBLP:conf/lics/Alfaro98,DBLP:conf/cav/AshokCDKM17,DBLP:conf/tacas/ButkovaWH17} (also known as steady-state or mean payoff values), cost-bounded properties~\cite{DBLP:conf/tacas/HartmannsJKQ18} (see \cref{sec:costbounded}), and support for multi-objective queries~\cite{DBLP:conf/tacas/ForejtKNPQ11,DBLP:conf/cav/QuatmannJK17} (see \cref{sec:multiobj}).

\subsection{Model Checking Methods}
\label{sec:methods}
In probabilistic model checking and arguably in verification in general, (sadly) there is no known ``one-size-fits-all'' solution.
Instead, the best tools and techniques depend heavily on the input model and the properties. 
\storm{}---as well as other model checkers---implements a variety of approaches that allow a knowledgeable user to pick the appropriate method as part of the input, and allows developers to extend and combine their favorite methods.
In particular we provide approaches based on solving (explicit) linear (in)equation systems, value-iteration variants on explicit or symbolic representations of (parts of the) model, policy iteration methods, methods using abstraction techniques and bisimulation minimization. 
We refer to \cref{sec:features} for some of \storm's distinguishing features for model checking, and \cref{sec:architecture_usage} for specifics on the technical realization.

%% file: contents/stormfeatures.tex
\section{\storm's Features}
\label{sec:features}
\begin{table*}
  \begin{center}
    \begin{tabular}{lc|cccc}
      \toprule
      feature & reference & DTMC & CTMC & MDP & MA \\
      \midrule
      sound/exact model checking & \cref{sec:soundexact} & \checkmark & $\checkmark^\ast$ & \checkmark & $\checkmark^\ast$ \\
      cost-bounded model checking & \cref{sec:costbounded} &\checkmark & \nocheckmark & \checkmark & \nocheckmark \\
      symbolic bisimulation minimization & \cref{sec:symbisim} & \checkmark & \checkmark & \checkmark & \checkmark \\
      game-based abstraction refinement & \cref{sec:gamebased} & \checkmark & \nocheckmark & \checkmark & \nocheckmark \\\midrule
      multi-objective model checking & \cref{sec:multiobj} & (\checkmark) & (\checkmark) & \checkmark & \checkmark \\
      high-level counterexamples & \cref{sec:counterex} & \checkmark & \nocheckmark & \checkmark & \nocheckmark \\
      parametric model checking & \cref{sec:parametric} & \checkmark & $\checkmark^\ast$ & \checkmark & $\checkmark^\ast$ \\\midrule
      partial observations & \cref{sec:pomdps} & $(\cdot)$ &$(\cdot)$ & \checkmark &  \nocheckmark \\
      dynamic fault trees & \cref{sec:dfts} & $(\cdot)$ & \checkmark & $(\cdot)$ & \checkmark \\\midrule
      
      permissive scheduler synthesis & \cite{DBLP:conf/tacas/Junges0DTK16} & (\checkmark) & (\nocheckmark) & \checkmark & \nocheckmark \\
      quantiles & \cite{HJKQ19} & \checkmark & \nocheckmark & \checkmark & \nocheckmark\\ 
      \bottomrule
    \end{tabular} \\[0.2cm]
    \begin{tabular}{rcl}
      $\checkmark^\ast$ & = & except for time-bounded reachability properties \\
      $(\cdot)$ & = & not meaningful
    \end{tabular}
  \end{center}
  \caption{Overview of distinguishing features of \storm{} and their applicability based on the model types.}
  \label{tab:features}
\end{table*}
In this section we  detail some of the outstanding features of \storm{} that go beyond conventional probabilistic model checking methods.
We give an overview in \cref{tab:features}.  

In particular, we have chosen four aspects that improve probabilistic model checking of standard properties such as reachability or expected rewards. These are reflected by the first four rows. 
\emph{Sound/exact} model checking reflects a collection of approaches that, compared to the classical numerical algorithms, provide stronger guarantees on the accuracy of the obtained results.
\emph{Cost-bounded} model checking, \emph{symbolic bisimulation minimization}, and \emph{game-based abstraction} reduce the size of the analyzed model in various ways to make probabilistic model checking more scalable.

Furthermore, we have selected three extensions that go beyond the classical variants of probabilistic model checking: 
we discuss how to extract \emph{counterexamples} using \storm{}, how to handle finding strategies that satisfy multiple properties simultaneously using \emph{multi-objective} model checking, 
we discuss \emph{parametric} models in which probabilities are not fixed constants but rather unknown symbols.

Finally, we discuss tailored model checking methods for \emph{POMDPs} and \emph{dynamic fault trees}.
We stress that the modular structure of \storm{} enables these approaches to easily reuse the regular model checking methods and the other methods outlined in this section.

\subsection{Exact and Sound Model Checking}
\label{sec:soundexact}
Several works~\cite{DBLP:conf/ddecs/WimmerKHB08,DBLP:conf/gi/Wimmer11,DBLP:conf/rp/HaddadM14,DBLP:conf/fmcad/BauerMCS017} observed that the numerical methods applied by probabilistic model checkers are prone to numerical errors.
This has mostly two reasons.
First, the floating point data types used by the tools are inherently imprecise.
For example, representing the probability $\frac{1}{10}$ using IEEE 754 compliant double precision introduces an error of $5 \cdot 10^{-18}$.
In the presence of numerical algorithms, these errors accumulate and may lead to incorrect results.
An alternative to the above is to employ \emph{rational arithmetic}.
That is, by representing probabilities (and costs) in the model and also the results as rational numbers, models may be analyzed without introducing any numerical errors.
\storm{} implements these ideas and allows for the exact solution of many properties.
However, efficient approaches for floating point arithmetic such as value iteration become inefficient when using rational numbers, as the representation of the latter grow very large.
\storm{} offers two tailored techniques to solve systems of (in)equations using rational arithmetic.
The first is based on policy iteration and Gaussian elimination and the second on a recent technique called \emph{rational search}~\cite{DBLP:conf/fmcad/BauerMCS017}.
The idea of the latter is to use an (imprecise) approximation of the exact solution and then sharpen this to a precise rational solution using the Kwek-Mehlhorn algorithm~\cite{DBLP:journals/ipl/KwekM03}.
If a straightforward check then returns that the sharpened values constitute an actual solution, the technique can return it.
Otherwise, the precision of the imprecise underlying solver is increased and the loop is restarted.

Secondly, the numerical algorithms sometimes themselves are strictly speaking unsound.
For example, standard value iteration for computing reachability probabilities  approximates the solution in the limit, but the termination criterion implemented by most tools does not guarantee that the obtained result is differing by at most the given precision $\epsilon$ from the actual solution.
One way to combat these problems is to approach the solution from both directions, a technique referred to as \emph{interval iteration}~\cite{DBLP:conf/atva/BrazdilCCFKKPU14,DBLP:conf/rp/HaddadM14,DBLP:conf/cav/Baier0L0W17}.
\storm{} implements the latter and additionally the more recent \emph{sound value iteration}~\cite{DBLP:conf/cav/QuatmannK18} and \emph{optimistic value iteration}~\cite{DBLP:conf/cav/HartmannsK20}.
Numerical errors aside\footnote{The implementation of these methods still uses finite precision floating point arithmetic.}, these methods ensure a correct result within a user-defined accuracy and come with a small time penalty as shown in~\cref{sec:evaluation}.

\subsection{Cost-bounded Reachability}
\label{sec:costbounded}
A typical application for Markov models is to analyze the probability to, e.g., reach a goal state before some resource like time or energy is depleted. 
Another typical application is to analyze the expected time before a number of tasks have been fulfilled. 
Both instances can be generalized to cost-bounded reachability. 
In cost-bounded reachability one is interested in the behavior of the system that does not violate the bounds on the resources.  
The classical approach to analyze cost-bounded reachability is to model this behavior in the model description by keeping track of the resources explicitly and then rely on standard reachability queries~\cite{DBLP:conf/formats/AndovaHK03}. 
That is, the states of the model keep track of the consumed resources, and the reachability query asks, e.g., what the probability is that one of the target states is reached in which the resource bounds are not violated.
The downside is that the model grows with these bounds. 

\storm{} alternatively allows modeling the (non-neg\-a\-tive) costs of actions or states in the modeling language. 
These costs are attached in the model, and then one may analyze cost-bounded reachability with the adequate query. 
The clear advantage of this approach is that the resources are not encoded in the state space which keeps the model much smaller. 
Rather, \storm{} does a series of model-checking calls on the much smaller model~\cite{DBLP:conf/tacas/HartmannsJKQ18,HJKQ19}, generalizing ideas from~\cite{DBLP:conf/setta/HahnH16,DBLP:journals/sttt/KleinBCDDKMM18} to multiple cost dimensions. 
The reduced memory footprint allows to handle much larger models, and often the reduced memory consumption also yields faster verification times. 

Cost-bounded reachability is closely related to quantile properties~\cite{DBLP:journals/sttt/KleinBCDDKMM18,HJKQ19}, 
where one fixes a desired reachability probability and asks how many resources have to be invested in order to achieve this probability.

\subsection{Symbolic Bisimulation Minimization}
\label{sec:symbisim}
A typical approach to alleviate the state space explosion is to represent the state space \emph{symbolically}.
%For qualitative systems, bounded model checking (BMC)~\cite{DBLP:journals/ac/BiereCCSZ03} or IC3~\cite{DBLP:conf/vmcai/Bradley11} (sometimes also referred to as property-driven reachability or PDR for short) are successful techniques that represent the state space using Boolean variables and logical formulae.
%Another route is the use of \acfp{bdd} for the representation of systems, an approach that in the qualitative case is mostly applied in the verification of hardware circuits.
In the probabilistic setting, employing variants of \emph{decision diagrams} (DDs) such as multi-terminal binary DDs (MTBDDs) or Multi-valued DDs (MDDs) is the most widely used approach to deal with large state spaces~\cite{DBLP:conf/icalp/BaierCHKR97}.
They are a graph-based data structure that can exploit structure and symmetry in the underlying model to represent gigantic models very compactly. 

A different angle to approach the problem is abstraction.
Here, the idea is to remove details from the model that are unnecessary for the desired analysis.
A well-studied technique is \emph{bisimulation minimization}. 
Its core idea is that states with equivalent behavior (in some suitable sense) can be merged to obtain a quotient model that preserves the properties of the original input.
Then, the (potentially much smaller) quotient can be analyzed instead.
Bisimulation minimization was shown to yield substantial reductions in the case that models are represented explicitly (for instance in terms of a probability matrix)~\cite{DBLP:conf/tacas/KatoenKZJ07}.

\storm{} allows to combine a symbolic representation with bisimulation minimization,
thereby extending previous work~\cite{DBLP:conf/gi/Wimmer11,DBLP:journals/sttt/DijkP18}. 
We extended the approach to deal with nondeterministic models, which makes it available on all four model types supported by \storm{} (see \cref{sec:modeltypes}).
This combination leads to significant reductions in memory and time consumption for a variety of models, and enables the analysis of models that are otherwise out of reach~\cite{DBLP:phd/dnb/Hensel18}.
The resulting quotient model is often small enough to be represented explicitly which enables a wide range of efficient analysis methods.

\subsection{Game-Based Abstraction-Refinement}
\label{sec:gamebased}
Even though bisimulation minimization effectively helps reducing the model, it has two major drawbacks.
First, it is not guided by the concrete analysis that is to be performed.
The quotient model may be much too fine for the analysis of a given property as it preserves a whole \emph{class} of properties.
Secondly, with few exceptions \cite{DBLP:conf/vmcai/DehnertKP13}, the algorithms to compute the bisimulation quotient require the entire state space and transitions to be available.
If the model is very large or even infinite, the algorithms fail to produce a quotient even if the quotient is very small.

Game-based abstraction \cite{DBLP:conf/qest/KwiatkowskaNP06} addresses these two challenges.
It is based on two fundamental ideas.
The first is that states are merged much more aggressively than in bisimulation minimization.
That is, they may be collapsed even if they have distinguishable behavior.
The behavior of the original model is over-approximated by the abstraction and the latter can therefore be used to obtain sound bounds for the measures on the former.
Note that the abstraction contains two sources of nondeterminism: the one present in the original model and the nondeterminism that is introduced by the abstraction process.
Merging these sources of nondeterminism results in very loose and unsatisfactory bounds on the target values.
The second idea therefore is to keep the two kinds of nondeterminism apart.
This gives rise to a stochastic game~\cite{DBLP:conf/dimacs/Condon90} whose solution gives lower and upper bounds on both minimal and maximal probabilities in the original model.

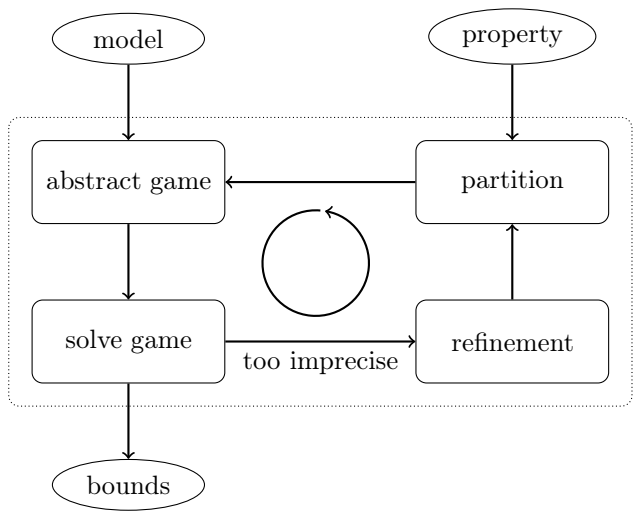
\begin{figure}
  \begin{center}
    \begin{tikzpicture}
      \tikzset{input/.style={draw, ellipse, minimum width=2cm}}
      \tikzset{processingstep/.style={draw, rectangle, rounded corners, align=center, text width=2.3cm, minimum height=1.1cm}}
      \tikzset{output/.style={input, minimum width=2cm}}
  	  \tikzset{controlflow/.style={->,thick}}

      \node[processingstep] (abstractgame) at (0,0) {abstract game};
      \node[processingstep, right=2.5cm of abstractgame] (partition) {partition};
      \node[processingstep, below=1cm of abstractgame] (solvegame) {solve game};
      \node[processingstep, below=1cm of partition] (refinement) {refinement};
    
      % inputs
      \node[input, above=1cm of abstractgame] (model) {model};
      \node[input, above=1cm of partition] (property) {property};
    
      % outputs
      \node[output, below=1cm of solvegame] (bounds) {bounds};
    
      % big box
      \draw[densely dotted, rounded corners] ($ (abstractgame.north west) + (0.3, -0.3) $) (abstractgame.north west) +(-0.3, 0.3) rectangle ($ (refinement.south east) + (0.3, -0.3) $);
    
      % arrows
      \draw[controlflow] (abstractgame.south) to (solvegame.north);
      \draw[controlflow] (solvegame.south) to (bounds.north);
      \draw[controlflow] (refinement.north) to (partition.south);
      \draw[controlflow] (solvegame.east) to node[auto, swap, ->] {too imprecise} (refinement.west);
      \draw[controlflow] (partition.west) to (abstractgame.east);

      \draw[controlflow] (model.south) to (abstractgame.north);
      \draw[controlflow] (property.south) to (partition.north);
    			
      \draw[controlflow] (abstractgame.south east) +(1.25cm, 0.18cm) arc	(85:440:0.7);
    \end{tikzpicture}
  \end{center}
  \caption{Overview of Abstraction-Refinement using games.}
  \label{fig:abstraction_refinement}
\end{figure}

\storm{} implements a game-based abstraction-re\-fine\-ment loop based on the ideas in \cite{DBLP:phd/de/Wachter2011}.
The loop is illustrated in \cref{fig:abstraction_refinement}.
As a first step, the abstract game is derived from the model and the current partitioning of the states, which is initially induced by the given property.
If the bounds obtained by the analysis of the game are precise enough, they can be returned.
Otherwise, the abstraction is refined by splitting the partition in a suitable way and the process is repeated.
To enable the analysis of gigantic or even infinite models, the abstraction is extracted directly from the high-level model description (given in terms of a \prismlang{} or \jani{} model).
This extraction is achieved by the formulation as a (series of) satisfiability problem(s), which are dispatched to an off-the-shelf solver.
While this has the aforementioned advantages, it is often the computationally most expensive part of the overall procedure.
To combat this, \storm{} implements several optimizations outlined in~\cite[Ch.\ 6]{DBLP:phd/dnb/Hensel18}.

\subsection{Multi-Objective Model Checking}
\label{sec:multiobj}
Initially, the focus in many probabilistic model checkers was mostly on computing the probability that a certain event happens.
However, probabilistic model checking can provide meaningful data beyond the probability to reach some state, such as the optimal strategies for MDPs, i.e., functions that describe how to resolve the nondeterminism in an MDP such that the induced behavior satisfies a given property.

However, if a strategy should satisfy multiple properties, standard model checking techniques do not suffice. Consider two properties limiting time and energy usage. Standard techniques would independently compute two strategies, one optimizing time, the other optimizing energy consumption. Both strategies might be wasting the other resource, thus violating the limits described in the matching combined property. 
Multi-objective model checking~\cite{DBLP:journals/lmcs/EtessamiKVY08,DBLP:conf/tacas/ForejtKNPQ11} helps in finding strategies that satisfy multiple properties at once, and can be used to clarify the trade-offs between various properties. 

Essentially, state-of-the-art multi-objective model checking boils down to a series of preprocessing steps on the model, and then either solving a linear program~\cite{DBLP:conf/tacas/ForejtKNPQ11} or iteratively applying standard model checking techniques~\cite{DBLP:conf/atva/ForejtKP12}. 
\storm{} supports multi-objective model checking on MDPs, and in addition on MAs \cite{DBLP:conf/cav/QuatmannJK17} under general as well as more restricted strategies~\cite{DBLP:conf/tacas/DelgrangeKQR20}. 
Furthermore, it allows for a more flexible combination of various properties, including properties with (multiple) cost-bounds~\cite{DBLP:conf/tacas/HartmannsJKQ18,HJKQ19}, and incorporates some particularly efficient preprocessing steps. 

\subsection{Synthesis of High-Level Counterexamples}
\label{sec:counterex}
%.
Besides the computation of a single strategy, the synthesis of counterexamples and/or of sets of strategies that all satisfy or violate a given property has gained some attraction. Here, we discuss counterexamples, but similar ideas have been used for so-called permissive strategies~\cite{DBLP:journals/corr/DragerFK0U15} as implemented in \storm{} using~\cite{DBLP:conf/tacas/Junges0DTK16}.

Suppose that a system reaches a bad state with a probability above some threshold. To locate the reason for this behavior, it is helpful to obtain the part of the system that leads to this behavior, by means of a counterexample. 
Counterexamples try capturing the essence of the failed verification attempt and help the user of the model checker---being a human or another algorithm---to revise the system or its model accordingly.
In the non-probabilistic setting, a counterexample may be represented as \emph{one} offending run of the system.
However, such a representation is not necessarily possible in the probabilistic setting as there may be infinitely many paths that contribute to the overall probability mass reaching the bad state~\cite{DBLP:journals/tse/HanKD09}.
A single run ending in a bad state is therefore typically insufficient as a counterexample.
While it is possible to consider sets of paths for probabilistic safety properties, the resulting counterexamples are large and hard to comprehend. 
Alternatively, counterexamples can be computed as sub-Markov models~\cite{DBLP:conf/sfm/AbrahamBDJKW14,DBLP:journals/tocl/ChadhaV10}.

Rather than considering counterexamples at the state-space level, \storm{} computes counterexamples in terms of the high-level model specification using the ideas of \cite{DBLP:conf/qest/WimmerJVAKB13}.
More concretely, given a \jani{} (or \prismlang{}) model that violates a safety property, \storm{} computes the smallest portion of the \jani{} code that already witnesses the violation based on the method proposed in \cite{DBLP:conf/atva/DehnertJWAK14}. 
It does so by a guided exploration of all candidate sub-models.
Ultimately, the smallest sub-model highlights the core of the problem. 
It does so at the abstraction level of the user.
High-level counterexamples are thus a valuable as diagnostic feedback to tool users (by humans).
Recent work has illustrated that these examples can be effectively used in a counterexample-guided inductive synthesis approach of finite Markov chains~\cite{DBLP:conf/fm/CeskaHJK19}.

\subsection{Parametric Model Checking}
\label{sec:parametric}
Naturally, the model checking result of Markov models crucially depends on the transition probabilities. 
Often, these probabilities are approximations based on data or reflect configurable parts of a modeled system.
To represent the uncertainty about the probabilities, parametric Markov models have been first considered in \cite{DBLP:conf/ictac/Daws04,DBLP:journals/fac/LanotteMT07}.
In parametric Markov models, the probabilities are symbolic expressions rather than concrete values.
For any valuation of the parameters, replacing the parameters in a parametric Markov model yields an \emph{instantiated} parameter-free Markov model.

There are many interesting questions that one can ask revolving around parametric systems.
The simplest is \emph{feasibility}, i.e., whether there exist a valuation such that the instantiated Markov model satisfies a property. 
More advanced is \emph{parameter space partitioning} where the goal is to decompose the parameter space into regions in which a predefined property is either satisfied or violated.
Such a decomposition indicates for \emph{most} parameter valuations whether they lead to a system that satisfies the given property.
An alternative question is to find the \emph{solution function}, i.e., a function in closed-form that gives the model checking result of the instantiated Markov model in terms of the parameter values. 
Already the feasibility problem is ETR-complete, that is, it is asymptotically as hard as finding the root of a multivariate polynomial~\cite{DBLP:conf/concur/WinklerJPK19}.

\storm{} supports the construction and analysis of parametric Markov models. 
Besides handling models and supporting efficient instantiation of parametric models, 
\storm{} provides three methods to perform parameter synthesis.
The first is based on computing the aforementioned solution function through state elimination~\cite{DBLP:conf/ictac/Daws04,DBLP:journals/sttt/HahnHZ11} that can also be seen as Gaussian elimination. 
This basic algorithm is improved by heuristics that order the operations, and a representation of the rational functions that allows for faster operations~\cite{DBLP:conf/cav/DehnertJJCVBKA15}.
The second method, referred to as \emph{parameter lifting}, avoids computing a potentially large rational function and determines validity of a formula over a region of parameter valuations through a sound abstraction into a non-parametric system~\cite{DBLP:conf/atva/QuatmannD0JK16}.
The third method~\cite{DBLP:conf/atva/SpelJK19} aims to analyze whether the solution function is monotonic in some parameter without actually computing the solution function, as the latter can be exponential in the number of parameters.
These and further methods are all used by the parameter synthesis tool \prophesy{}~\cite{DBLP:journals/corr/abs-1903-07993} which provides a playground for parameter synthesis approaches using \storm{} as a back-end.

\subsection{Partially Observable Markov Decision Processes}
\label{sec:pomdps}
\storm{} supports three methods for POMDP analysis:

First, \storm{} supports the \emph{verification} of (quantitative) reachability in POMDPs, e.g., to check whether for each policy resolving the nondeterminism based on the available observations, the probability to reach a bad state is less than $0.1$.
In general, this problem is undecidable~\cite{DBLP:journals/ai/MadaniHC03}.   
We consider an equivalent reformulation of the POMDP as an (infinite) belief MDP: 
Here, each state is a distribution over POMDP states. Such a belief MDP has additional properties that have been exploited to allow verification~\cite{DBLP:journals/ior/Lovejoy91,DBLP:journals/rts/Norman0Z17,DBLP:conf/ijcai/HorakBC18}.
\storm{} uses a combination of abstraction-and-refinement techniques to iteratively generate a finite abstract belief MDP that soundly approximates the extremal reachability probabilities in the POMDP~\cite{DBLP:journals/corr/abs-2007-00102}.

Often, POMDPs are analysed in settings where nondeterminism is controllable: The main interests is than in the dual of the verification problem: Find a policy such that the induced probability to reach a bad state is less than $0.1$. The problem remains undecidable.
A popular approach to overcome the hardness of the problem 
is to limit the policies, i.e., by putting a (small) a-priori bound on the memory of the policy~\cite{DBLP:conf/uai/MeuleauKKC99,DBLP:conf/uai/Hansen98,DBLP:conf/aaai/BraziunasB04,DBLP:journals/aamas/AmatoBZ10,DBLP:conf/nips/PajarinenP11,DBLP:conf/cdc/WintererJW0TK017}.
Such limits are especially reasonable when the nondeterminism is controllable, i.e., if a policy is to be synthesized. There are various cases in which small-memory policies deliver adequate performance. Additionally, these policies are small (and arguably simple) by construction.
 \storm{} translates POMDPs under observation-based policies with a fixed amount of memory to parametric DTMCs~\cite{DBLP:conf/uai/Junges0WQWK018}. Consider memoryless, observation-based policies: These policies map the current observation to a distribution over the available actions. We can encode all possible actions with the help of parameters. Finding values for these parameters then corresponds to finding an observation-based policy, and arguing over all parameters corresponds to arguing over all observation-based policies adhering to the memory limit.

Third, in POMDPs, even a qualitative variant of reachability is hard: 
In particular, to decide whether there exists a policy---resolving the nondeterminism based on the available observations---such that the probability to reach a bad state is $1$ is EXPTIME-complete~\cite{DBLP:conf/mfcs/ChatterjeeDH10}. 
\storm{} can compute small-memory policies via SAT-encodings~\cite{DBLP:conf/aaai/ChatterjeeCD16}, and finds more general policies by an incremental procedure~\cite{DBLP:journals/corr/abs-2007-00085}.

\subsection{Model Checking Dynamic Fault Trees}
\label{sec:dfts}
Fault trees~\cite{DBLP:journals/csr/RuijtersS15} are widely used in reliability engineering and model how component failures lead to failures of the complete system.
Dynamic fault trees (DFTs)~\cite{DBB90} extend (static) fault trees by dynamic gates.
DFTs more faithfully model systems by allowing order-dependent failures, functional dependencies and spare management.

Dynamic fault trees may be translated into corresponding Markov models~\cite{DBB90,DBLP:conf/dsn/BoudaliCS07} whose analysis yields common measures on dynamic fault trees, such as reliability and mean-time-to-failure.
The analysis of the corresponding Markov models also allows more complex measures, \eg, dealing with degraded modes~\cite{DBLP:journals/ress/GhadhabJKKV19}.
The essential step here is that \storm{} supports all these queries out-of-the-box.
Due to the modular architecture of \storm{} features such as parametric DFTs are supported off-the-shelf without dedicated implementation.

To drastically improve the analysis of DFTs, \storm{} contains a dedicated translation of such models into Markov models~\cite{DBLP:journals/tii/VolkJK18}.
To make the state-space generation as fast as possible, \storm{} utilizes the structure of the DFT, and constructs a Markov model that contains only the relevant behavior of the DFT.
Symmetries in the fault trees are exploited to further collapse the model with is then subject to regular model checking with \storm. 
As the state-space explosion might still be present during translation, \storm{} also supports a partial state-space generation for DFTs~\cite{DBLP:journals/tii/VolkJK18}.
This partial state space yields a sound abstraction, which may be model checked to obtain safe lower and upper bounds. 
The state space can be iteratively extended to obtain the desired precision of the analysis result.

%% file: contents/stormusage.tex
\section{Using \storm}

\storm{} is available as free and open software. Below, we give an overview how to use \storm. A detailed and up-to-date guide may be found on \storm{}'s website:
\begin{center}
\url{http://stormchecker.org}
\end{center}

\paragraph{Before you start.}
\storm{}  has to be configured and compiled on the target machine. This procedure automatically looks up various dependencies, and (optionally) adds them if they are not found on the system. 
While this configuration and compilation procedure offers some advantages, see \cref{sec:technicalities}, it is often cumbersome. Therefore, we recommend users which only want to experiment to rely on the \emph{docker containers}\footnote{Docker containers are a lightweight alternative to virtual machines. See \url{http://stormchecker.org/getting-started} for more details.} containing \storm{} with all the key dependencies, and all interfaces and extensions. 
One may start right away, at the cost of slightly reduced performance.

\paragraph{Model descriptions.}
\storm{} can be used with a variety of input languages including \jani{} and \prismlang{}. 
A complete up-to-date list and further resources can be found at \storm{}'s website\footnote{\url{http://stormchecker.org/documentation/languages}}.
For the sake of conciseness, we do not discuss the details of these languages here.

Below, we consider a \prismlang{} description of the Bounded Retransmission Protocol (brp)~\cite{DBLP:conf/types/HelminkSV93}. This and many other examples can be found in the \emph{Quantitative Verification Benchmark Set (QVBS)}\footnote{\url{http://qcomp.org/benchmarks}}~\cite{DBLP:conf/tacas/HartmannsKPQR19}.

\subsection{Command Line Interface}
The key way to interact with \storm{} is through its command line interface. The command line interface allows to specify the input model and properties, and after analysis reports on the requested results.
The command 
\[{\footnotesize\texttt{{\color{orange}storm} {\color{blue}-{}-prism} brp.pm {\color{blue}-{}-prop} brp.props }} \]
invokes \storm{} with a \prismlang{} description in \texttt{brp.pm}, and the properties listed in a file \texttt{brp.props}.
\storm{} will build the model and perform model checking on each property. 
For advanced users, the methods used for model checking can be flexibly yet simply set, e.g.,
\[{\footnotesize\texttt{{\color{orange}storm} ... {\color{blue}-{}-engine} hybrid {\color{blue}-{}-eqsolver} elimination }} \]
sets the engine to hybrid (see \cref{sec:engines}) and sets the linear equation solver to state elimination, see \cref{sec:solvers}. 
Experts may exploit the possibility to configure even details of the various procedures, e.g., the order in which state elimination is applied. 

\subsection{C++ Extensions}
To be able to flexibly use the internal data structures of \storm{}, one may build an own tool using \storm{} as a library. This approach is also taken by the \storm{} command-line interface, as well as other extensions shipped and tightly bound to \storm, such as the analysis of DFTs outlined in \cref{sec:dfts}.
This approach is the most flexible and powerful way of using \storm, but also requires most effort.
We illustrate model checking DTMCs with the sparse engine in \cref{fig:cppcodesnippet}.
The code parses a string and a property, builds a DTMC corresponding to the model, and applies model checking on the property to compute the corresponding probability for all states. The output is then created based on the model checking result of (some) initial state.
We provide a minimal working example to build your own \cpp{} tool based on \storm{} as a template repository\footnote{\url{http://stormchecker.org/api/starter-project}}.
\begin{figure*}
\input{contents/cppsnippet}
\caption{Using the \cpp{} interface (with \storm{} version 1.6.2). Please notice that we have omitted the necessary includes. An annotated version for the latest version is given in the starter project.}	
\label{fig:cppcodesnippet}
\end{figure*}

\subsection{Python Interface}
A much quicker way to flexibly interact with (a selection of) \storm's internal data structures is the \python{} API called \stormpy\footnote{Available from \url{http://stormchecker.org/stormpy} or via the python package index at \url{https://pypi.org/project/stormpy/}.}. 
We exemplify the ease of use in \cref{fig:pythonsnippet}. The code is equivalent to \cref{fig:cppcodesnippet}.
Using \python{} may induce some runtime penalty, but it enables a flexible access to the main functionality of \storm{}.
We stress that the code is powerful enough to drive also larger projects, e.g., the parameter synthesis tool \prophesy{} \cite{DBLP:conf/cav/DehnertJJCVBKA15} relies on \stormpy{}.
We provide a minimal working example to build your own \python{} tool based on \stormpy{} as a template repository\footnote{\url{http://stormchecker.org/stormpy/starter-project}}.

\begin{figure}
\input{contents/pythonsnippet}
\caption{Using \stormpy{} 1.6.2}	
\label{fig:pythonsnippet}
\end{figure}

\section{Architecture}
\label{sec:architecture_usage}
In this section, we report on some internal aspects of \storm. In particular, we aim to address how we realized performance and modularity. 
Naturally, we cannot go into the details of the various algorithms. Rather, we discuss some design choices that will help a user to feel more familiar with the code base. 

\subsection{Logical Structure}
The root directory of \storm{} contains---among others---sources and resources. The latter contains the logic for the configuration routines as well as various third-party dependencies. The sources are divided into various libraries and executables. The core functionality is found in the \texttt{storm} library. 
Inside that library, one finds data structures for the representation of matrices, models, expressions, modeling languages, as well as the model checking engines and solvers, which are discussed below.
Besides this library, there are libraries for parsing, handling parametric models, and handling various modeling formalism such as GSPNs and DFTs. All libraries depend on the core \texttt{storm} library.
Moreover, most libraries are accompanied by executables that provide adequate command line interfaces. 
\subsection{Models}
\storm{} features two different in-memory representations of Markov models.
First, it can use sparse matrices, an explicit representation form that uses memory roughly proportional to the number of transitions with non-zero probability.
Sparse matrices are suited for small and moderately-sized models and allow for fast operations also on models with irregular structure.
%In particular, \storm{} contains two ways of storing matrices: the regular sparse matrix which is stored in a compact form, or a more flexible sparse matrix that features more operations at the cost of slower analysis.
%\storm{} does not support dense matrices. 
Secondly, \storm{} can store models symbolically using MTBDD, cf.\ \cref{sec:symbisim}.
The MTBDDs are built from the model description directly. While it is possible to go from MTBDDs to the explicit representation, the other direction is not (efficiently) possible. 
While MTBDDs often store a model compactly, typical operations for the analysis of models yield a growth in the MTBDDs and are therefore often slow.
All models can be built representing the reachability probability with floating point arithmetic, exact rational numbers, or rational functions.

\begin{table*}[t]
		\begin{center}
			\begin{tabular}{rl}
				\toprule
				engine & supported features\\ 
				\cmidrule{1-2}
				\engine{sparse}, \engine{dd-to-sparse}, \engine{automatic} & all models and properties\\
				\engine{dd} & DTMC, MDP\\
				\engine{hybrid} & DTMC, CTMC, MDP, MA\\
				\engine{exploration}, \engine{abstraction-refinement} & reachability on DTMC and MDP\\
				\bottomrule     
			\end{tabular}
		\end{center}
		\caption{Overview of engines and supported features in \storm{}.}
		\label{tab:engines}
\end{table*}

\subsection{Model Checking Engines}
\label{sec:engines}
\storm's engines are built around the two model representations.
The \engine{sparse} engine exclusively uses the sparse matrix-based representation.
It first constructs the matrix representation of the state space by exploring the reachable state space specified in the modeling language, and then analyses the model using one of the many (standard, numerical) approaches, which are encapsulated as solvers (see below).
While the \engine{exploration} engine also uses sparse matrices, it uses ideas from reinforcement learning to avoid exploring all reachable states~\cite{DBLP:conf/atva/BrazdilCCFKKPU14}.
Instead, it proceeds in an ``on-the-fly'' manner and explores those parts of the system that appear to be most relevant to the verification task.

The next two engines use MTBDDs as their primary form of representation.
Except for the concrete in-memory representation, the \engine{dd} engine is the counterpart to the sparse engine in the sense that model building and verification is done on the very same representation and no translation takes place.
\storm's \engine{hybrid} engine tries to avoid the costly numerical operations on MTBDDs by transforming only parts of the system that are relevant for the considered property into a sparse matrix representation\footnote{This  approach corresponds to \prism's \engine{sparse} engine and is not to be confused with the latter's \engine{hybrid} engine, which is to be classified as ``more symbolical''.}. 

The \engine{dd-to-sparse} engine is similar, but performs the translation independent of the property. This can be useful when multiple properties are to be checked on the same model or when symbolic bisimulation minimization is applied. In the latter case, the quotient model will directly be constructed in a sparse matrix representation.

The \engine{abstraction-refinement} engine implements the technique described in \cref{sec:gamebased} and is able to compute bounds for both minimal and maximal reachability probabilities for (infinite) MDPs.

Given simple features of the input \prismlang{} or \jani{} model (such as the number of parallel automata or the average variable range), the \engine{automatic} engine automatically selects reasonable settings for \storm{}.
The current implementation uses a decision tree with 30 leaf nodes and a height of 7.
It has been generated with the tool \tool{scikit-learn}~\cite{scikit}  
using training data from experiments on the QComp benchmark set~\cite{DBLP:conf/tacas/HahnHHKKKPQRS19}.
To avoid over-fitting, the automatic choice only selects either
\begin{itemize}
\item the \engine{sparse} engine,
\item the \engine{sparse} engine with exact model checking and rational arithmetic (cf.~\cref{sec:soundexact}),
\item the \engine{hybrid} engine, or
\item the \engine{dd-to-sparse} engine with symbolic bisimulation minimization (cf.~\cref{sec:symbisim}).
\end{itemize}

\paragraph{Support for queries and model descriptions.}
\Cref{tab:engines} provides an overview of the models and queries supported by each engine.
The \engine{sparse} engine supports all model checking queries present in \storm{} and all DTMCs, CTMCs, MDPs, MAs, and POMDPs described in \prismlang{} or \jani. 
The engine can be paired with sound or exact model checking as in \cref{sec:soundexact}
However, exact arithmetic does not support time-bounded properties in CTMCs and MAs as these involve exponentials.
Many advanced features such as cost-bounded reachability and multi-objective model checking are only implemented in the \engine{sparse} engine.
The \engine{dd-to-sparse} engine can often make use of these implementations, as well.
The support within other engines is more limited.
The \engine{dd} engine does not support continuous time models (considered too slow) and POMDPs (typically sufficiently small).
The \engine{exploration} engine and the \engine{abstraction-refinement} engine are both limited to reachability queries on discrete-time models.
Moreover, some advanced features of the \jani{} language (indexed assignments, non-trivial system compositions) currently cannot be translated into DDs. 
The \engine{automatic} engine falls back to the sparse engine if the input model is not supported by the predicted configuration.

\subsection{Solvers}
\label{sec:solvers}
Probably the most outstanding trait of \storm's architecture is the concept of \emph{solvers}.
Ultimately, many tasks related to (probabilistic) verification revolve around solving subproblems.
For example, computing reachability probabilities or expected costs in a DTMC reduces to solving a system of linear equations.
Similarly, for an MDP a system of equations needs to be solved, with the difference that the equations are Bellman equations involving minima and maxima.
However, these are by no means the only kinds of problems appearing in probabilistic verification.
\begin{figure}
  \begin{center}
    \begin{tikzpicture}
      \tikzset{component/.style={align=center, rectangle, rounded corners, inner sep=3pt}}
      \tikzset{solvertype/.style={align=center, rectangle, rounded corners, minimum width=1.7cm, inner sep=3pt}}
      \tikzset{usesrelation/.style={dashed, ->}}
      
      \node[component] (sparse) {sparse};
      \node[component, below=1cm of sparse] (dd) {dd};    
      \node[component, below=1cm of dd] (hybrid) {hybrid};    
      \node[component, below=1cm of hybrid] (absref) {abstraction- \\ refinement};    
      \node[component, below=1cm of absref] (permissiveschedulers) {permissive \\ schedulers};    
      \node[component, below=1cm of permissiveschedulers] (counterexamples) {counterex.};    
      \node[component, below=1cm of counterexamples] (building) {model \\ building};    

      \node[solvertype, above right=0.075cm and 3cm of sparse.north east, anchor=north west] (linear) {linear \\ equations};
      \node[solvertype, below=1.717cm of linear] (bellman) {Bellman \\ equations};
      \node[solvertype, below=1.717cm of bellman] (stochastic) {stochastic \\ games};
      \node[solvertype, below=1.717cm of stochastic] (milp) {\milp};
      \node[solvertype, below=1.717cm of milp] (smt) {\smt};
    
      \draw[usesrelation] (sparse) to (linear);
      \draw[usesrelation] (sparse) to (bellman);
      \draw[usesrelation] (sparse) to (milp);

      \draw[usesrelation] (dd) to (linear);
      \draw[usesrelation] (dd) to (bellman);

      \draw[usesrelation] (hybrid) to (linear);
      \draw[usesrelation] (hybrid) to (bellman);
      \draw[usesrelation] (hybrid) to (milp);

      \draw[usesrelation] (absref) to (stochastic);
      \draw[usesrelation] (absref) to (smt);
    
      \draw[usesrelation] (permissiveschedulers) to (bellman);
      \draw[usesrelation] (permissiveschedulers) to (milp);
      \draw[usesrelation] (permissiveschedulers) to (smt);

      \draw[usesrelation] (counterexamples) to (bellman);
      \draw[usesrelation] (counterexamples) to (milp);
      \draw[usesrelation] (counterexamples) to (smt);

      \draw[usesrelation] (building) to (smt);
    
      \draw [decorate, thick, decoration={brace,amplitude=8pt}] ($(linear.north east)+(0.05,0)$) -- node[auto, xshift=16pt, yshift=18pt, rotate=-90] {solvers} ($(smt.south east)+(0.05,0)$);

      \coordinate (lowerleftengines) at ($(absref.south west)+(0.05,0)$);
      \draw [decorate, thick, decoration={brace,amplitude=8pt}] (lowerleftengines) -- node[auto, xshift=-16pt, yshift=18pt, rotate=90] {engines} (lowerleftengines |- sparse.north west);
    \end{tikzpicture}
  \end{center}
  \caption{Most important solvers used by \storm{}.}
  \label{fig:storm_solver_uses}
\end{figure}
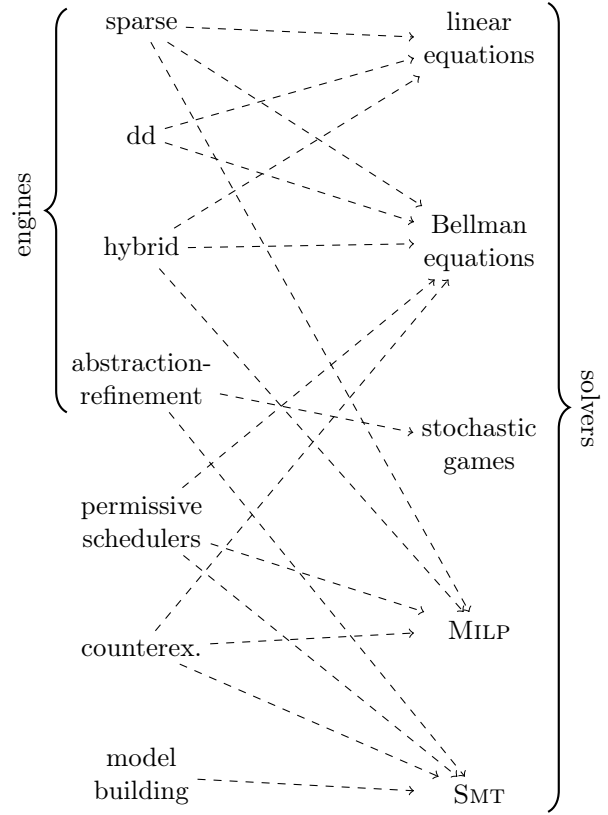

\begin{table*}
	\begin{center}
		\begin{tabular}{rl}
			\toprule
			solver type & available solvers \\
			\cmidrule{1-2}
			linear equations (sparse) & \eigen, \gmmxx, Gaussian elimination$^{*}$, native$^{*}$ \\
			linear equations (MTBDD) & \cudd, \sylvan \\
			Bellman equations (sparse) & \eigen, \gmmxx, native$^{*}$ \\
			Bellman equations (MTBDD) & \cudd, \sylvan \\
			stochastic games (sparse) & native$^{*}$ \\
			stochastic games (MTBDD) & \cudd, \sylvan \\
			(MI)LP & \gurobi, \glpk \\
			SMT & \zthree, \mathsat, \smtlib~\cite{BFT15} \\
			\bottomrule
		\end{tabular}
	\end{center}
	\caption{The solvers \storm{} provides out-of-the-box.}
	\label{tab:storm_solvers}
\end{table*}

\cref{fig:storm_solver_uses} illustrates some functionalities of \storm{} which have a dependency to one or more solvers.
For example, (explicit) model building employs \smt{} solving. 
As the initial states of symbolic models (\eg{} \prismlang{} or \jani) are given by the satisfying assignments of an expression, \storm{} uses \smt{} solvers to enumerate the possible initial states.
Similarly, the extraction of the abstract model from the symbolic model (as presented in \cref{sec:gamebased}) in the abstraction refinement engine crucially depends on enumerating satisfying assignments and therefore \smt{} solvers.
As yet another example, consider the synthesis of high-level counterexamples as in \cref{sec:counterex}.
Here, one of the offered techniques relies on the solution of a \milp{} while the other uses \smt{} solvers.

Two of the main goals in the development of \storm{} were the ability to exchange central building blocks (like solvers) and to benefit from (re)using high-performance implementations provided by other libraries.
It therefore offers abstract interfaces for the solver types mentioned above that are oblivious to the underlying implementation.
Offering these interfaces has several key advantages.
First, it provides easy and coherent access to the tasks commonly involved in probabilistic model checking.
Secondly, it enables the use of dedicated state-of-the-art high-performance libraries for the task at hand.
More specifically, as the performance characteristics of different backend solvers can vary drastically for the same input, this permits choosing the best solver for a given task.
Licensing problems are avoided, because implementations can be easily enabled and disabled, depending on whether or not the particular license fits the requirements.
Finally, implementing new solver functionality is easy and can be done without detailed knowledge of the global code base.
This flexibility allows to keep \storm{} up to date with new state-of-the-art solvers.

For each of the solver interfaces, several actual implementations exist.
For example, \storm{} currently has four implementations (each of them with a range of further options) of the linear equation solver interface for problems given as sparse matrices: one is based on \gmmxx{}, one on \eigen{}~\cite{eigenweb}, one uses its native internal data structures and algorithms for numerical algorithms and another one is based on Gaussian elimination~\cite{DBLP:conf/ictac/Daws04}.
\Cref{tab:storm_solvers} gives an overview over the currently available implementations. 
Here, all solvers that are purely implemented in terms of \storm's data structures and do not use libraries are marked with an asterisk to indicate that they are ``built-in''.

To realize the support for DD-based representations of systems, \storm{} relies on two different libraries: \cudd~\cite{cudd_website} and \sylvan~\cite{DBLP:phd/basesearch/Dijk16}.
While the former is very well established in the field, the latter is more recent and tries to make use of modern multi-core CPU architectures by parallelizing costly operations.
The parallelization comes at the price of more expensive bookkeeping and in general \cudd{} performs better if there are many operations on smaller DDs, while \sylvan{} is faster when fewer operations on larger DDs are involved.
%Given that modern CPUs tend to have more and more cores, \sylvan{} offers an interesting trade-off by trying to compensate for the slower operations on \acp{dd} (compared to explicit representations) by making efficient use of these resources.
\storm{} implements an abstraction layer on top of the two libraries that uses static polymorphism.
This way, it is possible to write code that is independent of the underlying library and does not incur runtime~costs.
%This also means that \storm{} can easily be extended with more \ac{dd} libraries as they become available.
%This sets it apart from \prism, which, because of its long-standing history, is somewhat coupled with \cudd.

\subsection{Technicalities}
\label{sec:technicalities}

By far the largest part (over 170,000 lines of code) of \storm{} is written in the \cpp{} programming language and extensively uses template meta-programming.
This has several positive and negative implications.
On the one hand, it serves the purpose of high performance for several reasons.
First, \cpp{} allows fine-grained control over implementation details like memory allocations.
Secondly, \cpp{} templates allow code to be heavily reused while maintaining performance as the static polymorphism enables type-dependent optimizations at compile-time.
Large parts of the code are written agnostic of the data type (floating point, rational number or even rational functions) and only the core parts are specialized based on the data type.
As this happens at compile-time, no runtime cost is incurred.
Finally, we observe that many high-performance solvers and data structure libraries that are well-suited for the context of (probabilistic) verification are written in \clanguage{} or \cpp{} (and also partially make use of template meta-programming), such as
\begin{itemize}
  \item SMT solvers (\zthree~\cite{DBLP:conf/tacas/MouraB08}, \mathsat~\cite{DBLP:conf/tacas/CimattiGSS13},  \smtrat~\cite{DBLP:conf/sat/CorziliusKJSA15}),
  \item LP solvers (\gurobi~\cite{gurobi_website}, \glpk{}\footnote{\url{https://www.gnu.org/software/glpk/}}),
  \item linear algebra libraries (\gmmxx{}\footnote{\url{http://getfem.org/gmm.html}}, \eigen{}~\cite{eigenweb}),
  \item DD libraries (\cudd~\cite{cudd_website}, \sylvan~\cite{DBLP:phd/basesearch/Dijk16}), and
  \item rational arithmetic libraries (\carl~\cite{DBLP:conf/sat/CorziliusKJSA15}, \gmp{}\footnote{\url{https://gmplib.org/}}).
\end{itemize}
Choosing \cpp{} as the language for \storm{} therefore allows easy and fast interfacing with these solvers.
On the other hand, the advantages come at a price.
Advanced templating patterns can be difficult to understand and increase compile-times significantly.

%% file: contents/cppsnippet.tex
{\scriptsize
\lstset{language=C++,
    style=FormattedNumber,
    basicstyle=\ttfamily,
    keywordstyle=\color{blue}\ttfamily,
    stringstyle=\color{red}\ttfamily,
    commentstyle=\color{green!50!black}\ttfamily,
    morecomment=[l][\color{magenta}]{\#}
    morekeywords={assert},
    otherkeywords={=,>,->},
%    morekeywords=[2]{check}, % doesn't work because there is another check inside the code
%    keywordstyle=[2]{\color{orange}},
    escapeinside={(>}{<)}
}
\begin{lstlisting}
#include ...

typedef storm::models::sparse::Dtmc<double> Dtmc;
typedef storm::modelchecker::SparseDtmcPrctlModelChecker<Dtmc> DtmcModelChecker;

bool (>{\color{orange}check}<)(std::string const& path_to_model, std::string const& property_string) {
    auto program = storm::parser::PrismParser::parse(path_to_model);
    // Code snippet assumes a Dtmc
    assert(program.getModelType() == storm::prism::Program::ModelType::DTMC);
    auto properties = storm::api::parsePropertiesForPrismProgram(property_string, program);
    auto formulae = storm::api::extractFormulasFromProperties(properties);
    auto model = storm::api::buildSparseModel<double>(program, formulae)->template as<Dtmc>();
    auto checker = std::make_shared<DtmcModelChecker>(*model);

    auto result = checker->check(storm::modelchecker::CheckTask<>(*(formulae[0]), true));
    assert(result->isExplicitQuantitativeCheckResult());
    // Use that we know that the model checker produces an explicit quantitative result
    auto quantRes = result->asExplicitQuantitativeCheckResult<double>();

    return quantRes[*model->getInitialStates().begin()] > 0.5;
}
\end{lstlisting}
}

%% file: contents/pythonsnippet.tex
{\scriptsize
\lstset{language=python,
    style=FormattedNumber,
    basicstyle=\ttfamily,
    keywordstyle=\color{blue}\ttfamily,
    stringstyle=\color{red}\ttfamily,
    commentstyle=\color{green}\ttfamily,
    morecomment=[l][\color{magenta}]{\#},
    morekeywords={as},
    otherkeywords={=,>},
    morekeywords=[2]{check},
    keywordstyle=[2]{\color{orange}},
}

\begin{lstlisting}
import stormpy as sp

def check(path_to_model, property_str):
  program = sp.parse_prism_program(path_to_model)
  props = sp.parse_properties(property_str, program)
  model = sp.build_model(program, props)
  result = sp.model_checking(model, props[0])
  return result.at(model.initial_states[0]) > 0.5
\end{lstlisting}
}

%% file: contents/evaluation.tex
\begin{table*}
	\centering
		\setlength{\tabcolsep}{3.5pt}
		\begin{tabular}{rrrrrrrr}
			\toprule
			& \multicolumn{1}{c}{\engine{sparse}}
			& \multicolumn{1}{c}{\engine{hybrid}}
			& \multicolumn{1}{c}{\engine{dd}}
			& \multicolumn{1}{c}{\engine{bisim}}
			& \multicolumn{1}{c}{\engine{sound}}
			& \multicolumn{1}{c}{\engine{exact}}
			& \multicolumn{1}{c}{\engine{automatic}}
			\\\cmidrule{1-8}
			\#solved     & 73	& 67	& 40	& 59	& 73	& 43	& 84\\
			\#not supp.  & 0	& 11	& 42	& 7	& 0	& 14	& 0\\
			\#time-outs  & 3	& 5	& 4	& 8	& 3	& 12	& 3\\
			\#mem-outs   & 16	& 11	& 8	& 20	& 18	& 27	& 7\\
			\#incorrect  & 4	& 2	& 2	& 2	& 2	& 0	& 2\\
			\hline
			\#fastest$_{+1\%}$ & 19	& 21	& 9	& 14	& 8	& 3	& 40\\
			\#fastest$_{+50\%}$ & 39	& 46	& 16	& 26	& 27	& 4	& 78\\
			\bottomrule
		\end{tabular}
		
		% A total of 87 instances could be solved.
		\caption{Outcomes of experiments on 96 benchmark instances.}
		\label{tab:results}
\end{table*}

\section{Evaluation}
\label{sec:evaluation}
This section contains an empirical evaluation of some key functionalities of \storm.
Furthermore, we recap results of QComp 2019~\cite{DBLP:conf/tacas/HahnHHKKKPQRS19} and QComp 2020~\cite{qcomp2020} to emphasize the competitiveness of \storm.
%\footnote{}

\subsection{Setup and Methodology}
\label{sec:setup_methodology}

We consider the set of 100 benchmark instances that were selected in QComp 2019 and 2020~\cite{DBLP:conf/tacas/HahnHHKKKPQRS19,qcomp2020}.
Each instance consists of a symbolic model description and a property specification from the \emph{Quantitative Verification Benchmark Set (QVBS)}~\cite{DBLP:conf/tacas/HartmannsKPQR19}.
If available, we consider model descriptions in the \prismlang{} language.
Otherwise, the model is build from the \jani{} description.
For a better comparison across \storm's engines, we did \emph{not} employ the techniques from \Cref{sec:dfts} to solve DFTs.
Since \storm{} has no native support for PTA, we used the tool \tool{moconv} (part of the \modesttoolset{}\footnote{\url{http://www.modestchecker.net/}}~\cite{DBLP:conf/tacas/HartmannsH14}) to translate PTAs into MDPs.
For four instances either \tool{moconv} did not support the PTA or \storm{} did not support the output of \tool{moconv}.
We therefore restrict our evaluation to the remaining 96 benchmark instances.

For each instance, the task is to solve the corresponding model checking query within a time limit of 30 minutes and a memory limit of 12 GB.
The results are compared to the reference results provided by the QVBS.
If the relative difference between these values is greater than $10^{-3}$, the result is considered \emph{incorrect}.
This setup coincides with the setup of QComp 2019.
All experiments were run on 4 cores of an Intel\textsuperscript{\textregistered} Xeon\textsuperscript{\textregistered} Platinum 8160 Processor.
We measure the wall-clock runtimes (including model building and model checking) for all experiments. 
Notice that this machine is more powerful than the QComp 2019 machine.

For our evaluation we consider \storm{} version 1.6.2 in \emph{seven different configurations} comprising
\begin{itemize}
\item the main engines of \storm{}: \engine{sparse}, \engine{hybrid}, and \engine{dd},
\item symbolic bisimulation (\engine{bisim}) with sparse quotient (\Cref{sec:symbisim}),
\item \engine{sound} and \engine{exact} model checking within the \engine{sparse} engine (\Cref{sec:soundexact}), and
\item the \engine{automatic} engine.
\end{itemize}
Whenever the invoked model checking method is sound (\ie{} provides precision guarantees), the precision of \storm{} is set to $10^{-3}$ (relative).
Otherwise, \storm{}'s default precision $10^{-6}$ (relative) is used.
We select \sylvan~\cite{DBLP:phd/basesearch/Dijk16} as DD-library, and set its memory limit to 4 GB.
We also consider a ``fastest'' configuration that takes the best result from the seven configurations, i.e., a configuration which runs all seven configurations and terminates whenever the fastest terminates (and further runs the seven configurations independently on different machines).

All benchmark files, log files and replication scripts are available at~\cite{replication_package}.
\subsection{Results}
\label{sec:results}
\begin{figure*}[t]
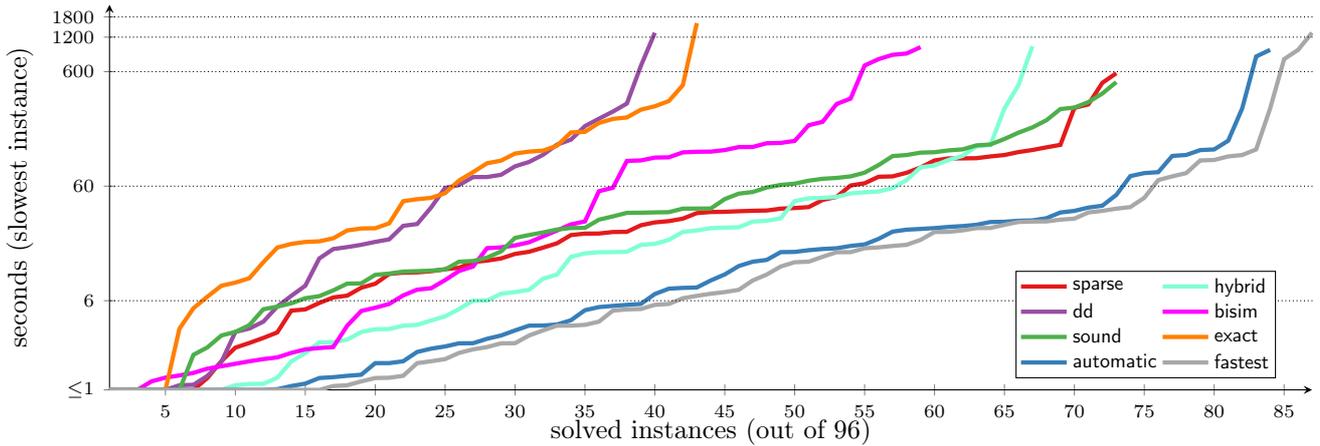

	\centering
	\setlength{\quantileplotwidth}{\linewidth}
	\renewcommand{\quantileplotlegendcols}{2}
	\quantileplot{contents/plotdata/quantile.csv}%
	{Storm.sparse,			Storm.hybrid,			Storm.dd,		Storm.ddbisim, Storm.sound, Storm.exact,	Storm.automatic,   Storm.best}%
	{\engine{sparse},	\engine{hybrid},	\engine{dd},	\engine{bisim}, \engine{sound}, \engine{exact}, 	\engine{automatic}, 	\engine{fastest}}%
	{96} % num of benchmarks in total
	\caption{Runtime comparison of \storm{}'s key features.}
	\label{fig:QuantilePlotStorm}
\end{figure*}
\Cref{tab:results} summarizes the outcomes of our experiments. The seven columns refer to the seven configurations as described above.
In the first row we indicate how many of the 96 considered instances were correctly solved for each configuration.
The subsequent rows indicate the number of not supported instances\footnote{Observe that \engine{sparse}, \engine{sound}, and \engine{automatic} support all queries. For details on the other configurations see \cref{sec:engines}.}, the number of times the time- or  memory limit was exceeded, respectively, and the number of incorrect results\footnote{The incorrect results are the consequence of imprecise floating-points or algorithms that do not guarantee sound results, see \Cref{sec:soundexact}.} that were obtained. Observe that these rows always sum to 96.
%MAs are not supported in the \engine{hybrid} engine does not support MAs and the \engine{dd} engine  The DD based engines do not support timed properties on CTMCs and MAs as these are considered to slow to be relevant,

For the ``fastest'' configuration, we obtain 87 solved instances and 0 incorrect results.
The next rows (after the horizontal line) show how often each configuration was either the fastest among the tested ones or only 1\% (50\%) slower than the fastest one, i.e., terminated within 101\% (150\%) of the fastest configuration.

We further compare the runtimes of the different engines and features in \Cref{fig:QuantilePlotStorm}.
The shown quantile plot expresses how many benchmark instances (measured on the x-axis) \emph{each} were solved in at most the time given on the y-axis.
In other words, the point $\tuple{x, y}$ is contained in the quantile plot for configuration \engine{c} if the \emph{maximal} runtime when using \engine{c} on the $x$ fastest solved instances (for \engine{c}) is $y$ seconds.
%In particular, this means that times are \emph{not} cumulative and do therefore \emph{not count the start-up times of tools repeatedly}.
Time- and memory-outs, incorrect results and unsupported experiments may skew the lines of the affected configurations as all these outcomes do \emph{not} count as solved.
Besides the seven considered configurations we also depict the runtime obtained by the \engine{fastest} engine or feature for each individual benchmark.

\begin{figure*}[p]
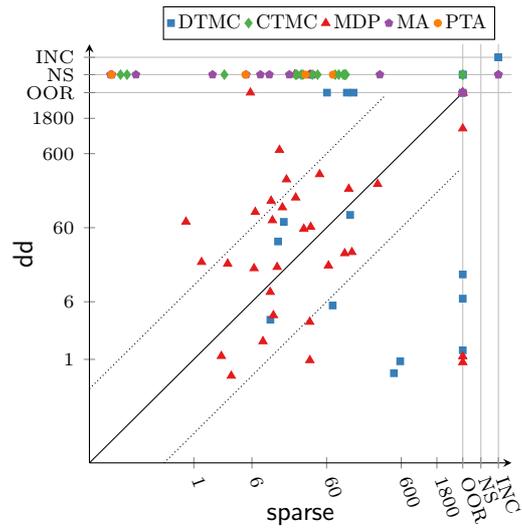
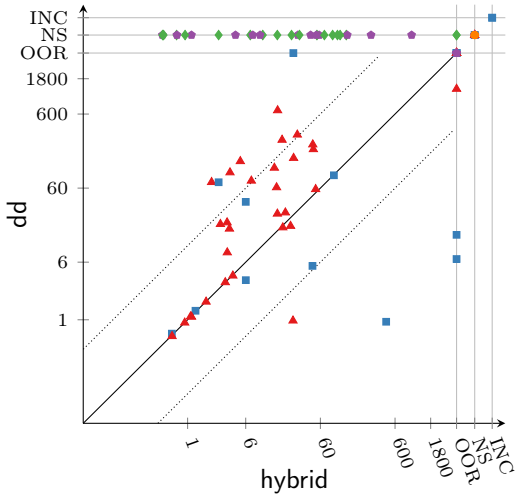
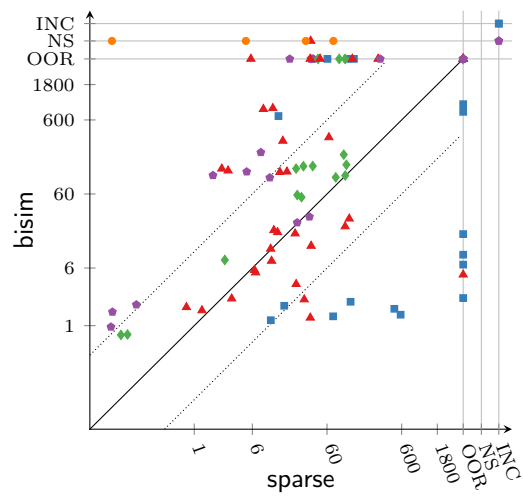
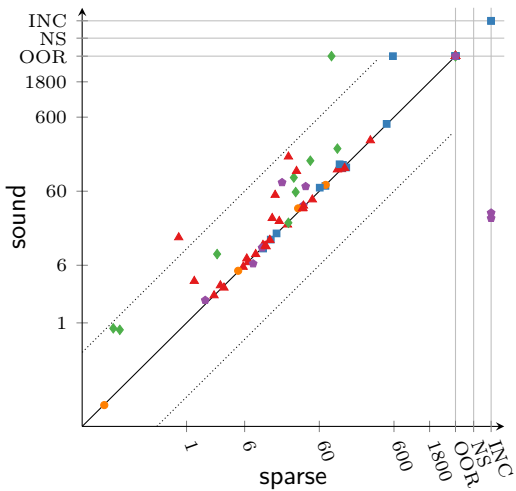
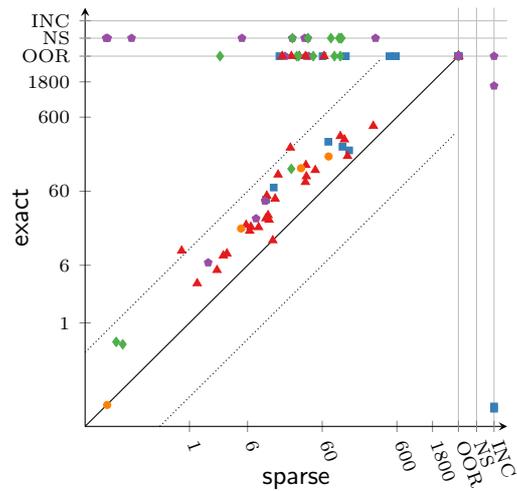

	\centering
	\setlength{\scatterplotsize}{0.41\linewidth}
	\subfigure[Comparison of \engine{sparse} and \engine{hybrid} engine]{
		\scatterplotstorm{contents/plotdata/scatter.csv}{Storm.sparse}{\engine{sparse}}{Storm.hybrid}{\engine{hybrid}}{true}
		\label{fig:ScatterStormsparsehybrid}
	}
	\subfigure[Comparison of \engine{sparse} and \engine{dd} engine]{
		\scatterplotstorm{contents/plotdata/scatter.csv}{Storm.sparse}{\engine{sparse}}{Storm.dd}{\engine{dd}}{true}
		\label{fig:ScatterStormsparsedd}
	}
	\\
	\subfigure[Comparison of \engine{hybrid} and \engine{dd} engine]{
		\scatterplotstorm{contents/plotdata/scatter.csv}{Storm.hybrid}{\engine{hybrid}}{Storm.dd}{\engine{dd}}{false}
		\label{fig:ScatterStormhybriddd}
	}
	\subfigure[Evaluation of symbolic bisimulation]{
		\scatterplotstorm{contents/plotdata/scatter.csv}{Storm.sparse}{\engine{sparse}}{Storm.ddbisim}{\engine{bisim}}{false}
		\label{fig:ScatterStormBisim}
	}
	\\
	\subfigure[Evaluation of sound model checking]{
		\scatterplotstorm{contents/plotdata/scatter.csv}{Storm.sparse}{\engine{sparse}}{Storm.sound}{\engine{sound}}{false}
		\label{fig:ScatterStormSound}
	}
	\subfigure[Evaluation of exact model checking]{
		\scatterplotstorm{contents/plotdata/scatter.csv}{Storm.sparse}{\engine{sparse}}{Storm.exact}{\engine{exact}}{false}
		\label{fig:ScatterStormExact}
	}
	\caption{Comparison of engines and features of \storm{}.}
	\label{fig:ScatterStorm}
\end{figure*}

\begin{figure*}[t]
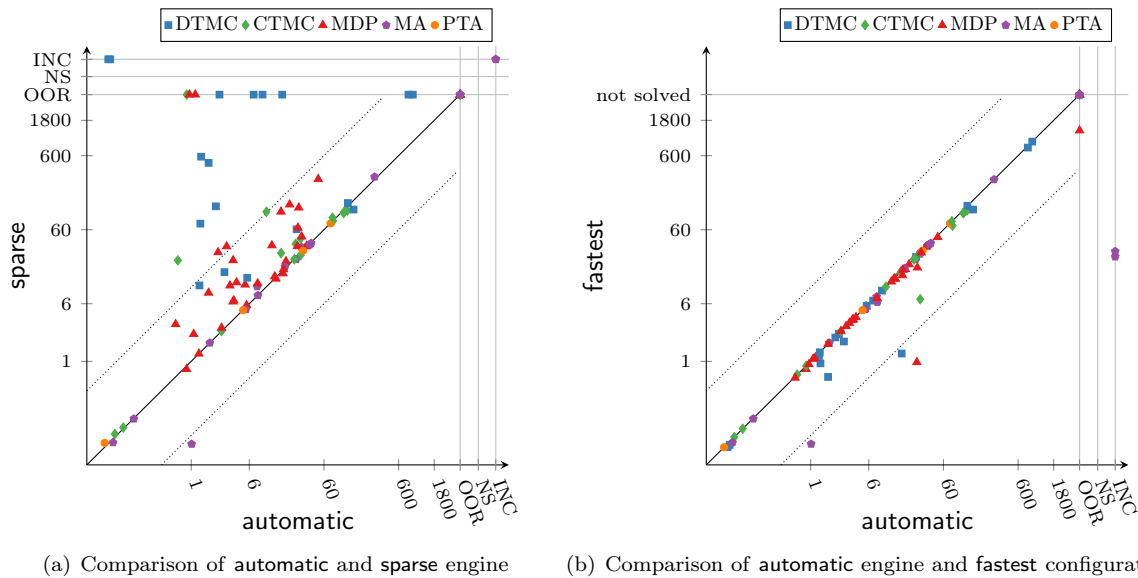

	\centering
	\setlength{\scatterplotsize}{0.41\linewidth}
	\subfigure[Comparison of \engine{automatic} and \engine{sparse} engine]{
		\scatterplotstorm{contents/plotdata/scatter.csv}{Storm.automatic}{\engine{automatic}}{Storm.sparse}{\engine{sparse}}{true}
		\label{fig:ScatterStormAutoSparse}
	}
	\subfigure[Comparison of \engine{automatic} engine and \engine{fastest} configuration]{
		\scatterplotstormfastesty{contents/plotdata/scatter.csv}{Storm.automatic}{\engine{automatic}}{best}{\engine{fastest}}{true}
		\label{fig:ScatterStormAutoFastest}
	}
	\caption{Comparison of engines and features of \storm{} (continued).}
	\label{fig:ScatterStormContinued}
\end{figure*}

Finally, we compare the configurations of \storm{} one-by-one and give the results in \Cref{fig:ScatterStorm,fig:ScatterStormContinued}.
Each point in the depicted scatter plots indicates the runtimes of the two compared configurations for one benchmark instance. The type (DTMC, CTMC, MDP, MA, or PTA) of the verification task is indicated by means of different marks.
The scatter plots use logarithmic scales on both axes and indicate speed-ups of $10$ by means of dotted lines.
If an experiment ran out of resources (time or memory), was not supported, or yielded an incorrect result, we draw the point on separate lines, labeled OOR, NS, and INC, respectively.
We compare the engines (\engine{sparse}, \engine{hybrid}, and \engine{dd}) with each other in \Cref{fig:ScatterStormsparsehybrid,fig:ScatterStormsparsedd,fig:ScatterStormhybriddd}.
Symbolic bisimulation, sound, and exact model checking are compared with the \engine{sparse} engine (the default of \storm) in \Cref{fig:ScatterStormBisim,fig:ScatterStormSound,fig:ScatterStormExact}.
For the comparison with sound model checking, we do not depict benchmark instances where the default method is already sound. 
\Cref{fig:ScatterStormAutoSparse,fig:ScatterStormAutoFastest} compare the \engine{automatic} engine with the \engine{sparse} engine and with the \engine{fastest} configuration, respectively.

More detailed results of our experiments can be found on~\url{http://stormchecker.org/benchmarks}.

\subsection{Discussion}
Comparing the three main engines of \storm{} (\engine{sparse}, \engine{hybrid}, and \engine{dd}), the \engine{sparse} engine was the most versatile engine during our experiments since it supports all 96 instances and successfully solved the majority (73) of them, outperforming the other two engines.
However, looking at \Cref{fig:QuantilePlotStorm} we see that the other engines are competitive.
The \engine{automatic} engine often manages to pick the ``right'' configuration for a given benchmark and thus almost matches the performance of the (notional) \engine{fastest} configuration.
As indicated in \Cref{fig:ScatterStorm}, several instances could only be solved using symbolic techniques based on the \engine{hybrid} or the \engine{dd} engine.
We emphasize that the benchmark selection can have a strong impact when comparing the engines of \storm{} because the symbolic engines are strongly reliant on the model structure.
Moreover, many benchmarks are not supported by the \engine{hybrid} and/or the \engine{dd} engine which skews the lines in \Cref{fig:QuantilePlotStorm}.

Symbolic bisimulation was extremely effective on models with a concise DD-based representation and a small bisimulation quotient.
The export into a sparse quotient allows \storm{} to make use of the versatility of the \engine{sparse} engine.

In \Cref{fig:ScatterStormSound} we see that the overhead for sound model checking is often negligible.
As mentioned above, we invoke classical model checking (such as value iteration) with the default precision parameters ($10^{-6}$ relative precision) whereas sound model checking  is invoked with the  actual precision requirements ($10^{-3}$ relative precision), yielding speed-ups for some instances.

Exact model checking is comparably costly. The use of exact (infinite precision) arithmetic induces increasingly larger number representations. Moreover, approximative, numerical solution methods cannot be applied.
However, on a few instances where numerical methods do not work well, exact model checking was superior to the remaining configurations.

\Cref{fig:ScatterStormAutoSparse,fig:ScatterStormAutoFastest} show that---for this benchmark set---the \engine{automatic} engine improved the runtime of the \engine{sparse} engine in many cases and that there were only a few instances where it was outperformed by the (notional) \engine{fastest} engine. 

\subsection{Summary from QComp 2019}
\label{sec:qcomp}

We briefly recap the results of QComp 2019, focusing on the performance evaluation.
For further details we refer to the competition report~\cite{DBLP:conf/tacas/HahnHHKKKPQRS19}.

The experimental setup of QComp 2019 (benchmark selection, precision requirements, time- and memory limits, \dots) coincides with our setup as detailed above, except that
\begin{itemize}
	\item a different machine was used, and
	\item \storm{} was considered in version 1.3.0.
\end{itemize}
Each tool was executed in two different modes: Once with default settings (which for \storm{} coincides with using the \engine{sparse} engine) and once with benchmark specific settings.
For the latter mode, the participants could provide a tailored tool invocation for each individual benchmark instance.
For \storm{} this was realized by empirically determining the fastest configuration for a given instance, where we considered the configurations \engine{sparse}, \engine{hybrid}, \engine{dd}, \engine{bisim}, \engine{sound}, and \engine{exact} (as above).

\begin{figure*}[t]
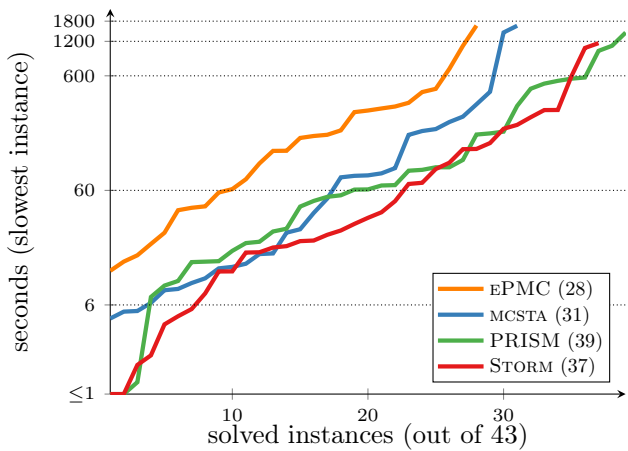
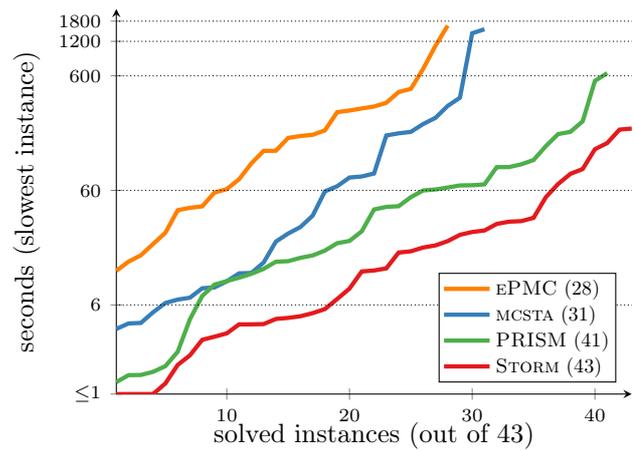
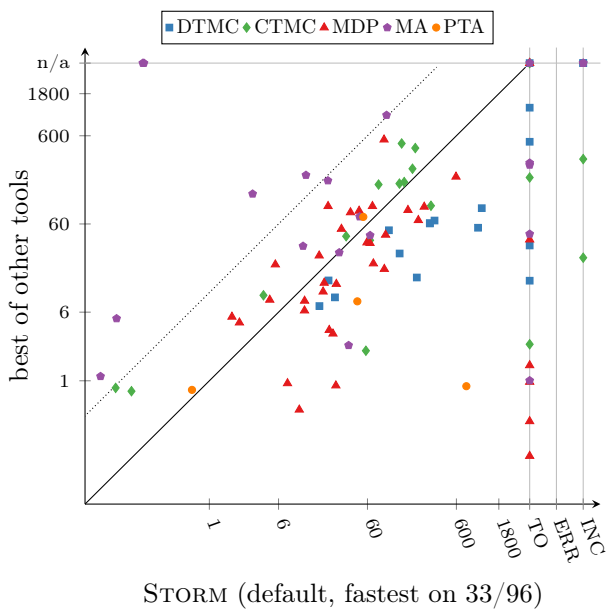
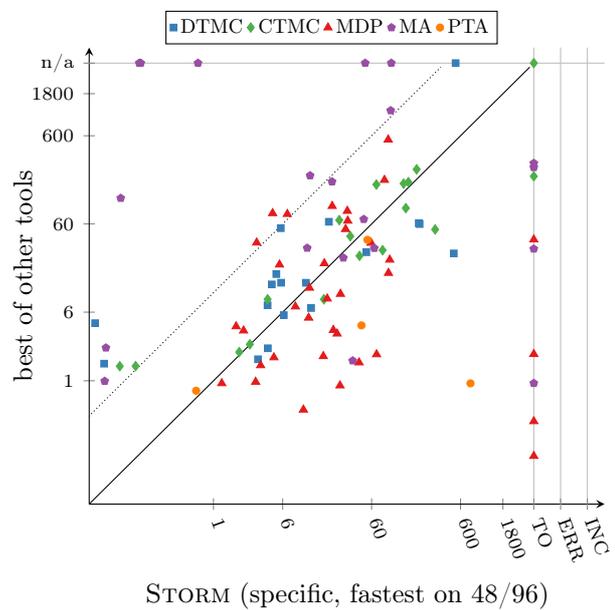

		\begin{center}
		\subfigure[Quantile plots for the general-purpose model checkers (default)]{
	\quantileplot{contents/plotdata/qcomp/generalpurpose.csv}{epmc,mcsta,PRISM,Storm}{\tool{ePMC} (28),\tool{mcsta} (31),\tool{PRISM} (39),\tool{Storm} (37)}{43}
	\label{fig:Qcomp:quantile:default}
	}
	\subfigure[Quantile plots for the general-purpose model checkers (specific). Repeats \cref{fig:QuantilePlotsTweaked}.]{
		\quantileplot{contents/plotdata/qcomp/specificgeneralpurpose.csv}{epmc,mcsta,PRISM,Storm}{\tool{ePMC} (28),\tool{mcsta} (31),\tool{PRISM} (41),\tool{Storm} (43)}{43}
	\label{fig:Qcomp:quantile:specific}
	}
	\\
	\subfigure[Runtime of Storm (default) compared with the best results.]{
	\scatterplot{contents/plotdata/qcomp/scatterStorm.csv}{Storm}{\tool{Storm} (default, fastest on 33/96)}{true}
	\label{fig:Qcomp:scatter:default}
	}
	\subfigure[Runtime of Storm (best config) compared with the best results.]{
	\scatterplot{contents/plotdata/qcomp/scatterspecificStorm.csv}{Storm}{\tool{Storm} (specific, fastest on 48/96)}{true}
	\label{fig:Qcomp:scatter:specific}
	}	
	\end{center}
	\caption{The performance of \storm{} compared with other state-of-the-art model checkers. All figures are taken from~\cite{DBLP:conf/tacas/HahnHHKKKPQRS19}$^{*}$}
	\footnotesize{$^{*}$: Licensed under~Creative Commons Attribution 4.0 International License: \url{http://creativecommons.org/licenses/by/4.0/}.}
	\label{fig:Qcomp}
\end{figure*}

\Cref{fig:Qcomp} depicts the performance results of QComp 2019 that are relevant for \storm{}.
The quantile plots in \Cref{fig:Qcomp:quantile:default,fig:Qcomp:quantile:specific} compare \storm{} with the other participating general-purpose probabilistic model checkers \epmc~\cite{DBLP:conf/fm/HahnLSTZ14}, \mcsta~\cite{DBLP:conf/atva/HartmannsH15}, and \prism~\cite{DBLP:conf/cav/KwiatkowskaNP11} using the default and specific modes, respectively.
\storm{} supported 96 of the 100 considered benchmark instances, whereas \epmc, \prism, and \mcsta{} supported 63, 58, and 86 instances, respectively. 
For the quantile plots only the 43 instances that were supported by all 4 tools were taken into account.
In particular, all benchmarks are given in \prismlang{} language since \prism{} does not support \jani.
The scatter plots in \Cref{fig:Qcomp:scatter:default,fig:Qcomp:scatter:specific} compare \storm{} with the best of the other 8 participating tools.
A point above the solid diagonal line indicates that on the corresponding instance, \storm{} was the fastest tool among all participants.

Considering the results for the default mode in \Cref{fig:Qcomp:quantile:default}, \storm{} is the strongest competitor of the other three tools. However, the performance results of \storm{} and \prism{} are very close to each other.
For instance-specific invocations (\Cref{fig:Qcomp:quantile:specific}), \storm{} clearly outperformed all its competitors.
The scatter plots show that \storm{} performed best among all tools for $1/3$ of the supported benchmarks in default mode and $1/2$ of the supported benchmarks in specific mode.

\subsection{Outlook to QComp 2020}
Since QComp 2019 further progress of the participating tools has been made. 
For example, new and efficient model-checking techniques for MDPs and MAs have been implemented in \mcsta~\cite{DBLP:conf/qest/ButkovaHH19,DBLP:conf/cav/HartmannsK20}.
QComp 2020~\cite{qcomp2020} captures some of these changes and gives a special emphasis to the correctness of the results produced by the tools.
In contrast to the 2019 edition, the performance evaluation is divided in six tracks. The tracks consider the same benchmark set but impose different correctness requirements ranging from \emph{exact} results to \emph{often $\varepsilon$-correct} results.
Among all nine participants, \storm{} has been the only tool that implements supporting algorithms for \emph{all} tracks and has proven competitiveness in each of them.
More details to QComp 2020 can be found in its competition report~\cite{qcomp2020}.

We remark that both QComp 2019 and QComp 2020 necessarily only provide a snapshot of the tool landscape at the time of the evaluation.
A repetition of the evaluation of QComp with newer tool versions can yield different results.

%% file: contents/conclusion.tex
\section{Conclusion}
\label{sec:conclusion}
This paper presented the state-of-the-art probabilistic model checker \storm{}.
We have discussed its main distinguishing features, and described how it can be used for rapid prototyping of new algorithms and tools.
Key aspects of \storm{} are its modularity, its accessibility through a Python interface, its various modeling formalisms, as well as the functionalities that go beyond the standard probabilistic model-checking algorithms.
We believe that its modularity, careful crafting of the most time-consuming operations, and our experience with earlier in-house developed model checkers, have led to a tool that is competitive to existing probabilistic model checkers.
\storm{} provides an effective and efficient platform for future-proof developments in probabilistic model checking.
It is open access and publicly available from \url{http://stormchecker.org}.
A major challenges will be to keep up with the rapid progress in the field.
This does not only involve the implementation of new algorithms, but also involves constantly revising existing code fragments.

%% file: contents/acknowledgements.tex
\paragraph*{Acknowledgements.} 

We would like to thank all additional contributors to \storm{}, in particular 
Philipp Berger, David Korzeniewski, Jip Spel,
as well as Dimitri Bohlender, Alexander Bork, Harold Bruintjes, Michael Deutschen, Timo Gross, Thomas Heinemann, Thomas Henn,
Tom Janson, Jan Karuc, Joachim Klein, Gereon Kremer, Sascha V.\ Kurowski, Manuel S.\ Weiand, and Lukas Westhofen. 
We would like to thank Tom van Dijk for his support regarding \sylvan. 

Furthermore, we would like to thank the users for being interested, and for the many kind bug-reports that helped us improving the tool. 

This work has been supported by the ERC Advanced Grant 787914 (FRAPPANT) and the DFG RTG 2236 `UnRAVeL'.
S.~Junges would like to acknowledge funding from  NSF grants 1545126 (VeHICaL) and 1646208, the DARPA Assured Autonomy program,  Berkeley Deep Drive, and by Toyota under the iCyPhy center.

%% file: main.bbl
\begin{thebibliography}{100}
\providecommand{\url}[1]{{#1}}
\providecommand{\urlprefix}{URL }
\expandafter\ifx\csname urlstyle\endcsname\relax
  \providecommand{\doi}[1]{DOI~\discretionary{}{}{}#1}\else
  \providecommand{\doi}{DOI~\discretionary{}{}{}\begingroup
  \urlstyle{rm}\Url}\fi

\bibitem{DBLP:conf/sfm/AbrahamBDJKW14}
{\'{A}}brah{\'{a}}m, E., Becker, B., Dehnert, C., Jansen, N., Katoen, J.P.,
  Wimmer, R.: Counterexample generation for discrete-time {M}arkov models: An
  introductory survey.
\newblock In: {SFM}, \emph{{LNCS}}, vol. 8483, pp. 65--121. Springer (2014)

\bibitem{DBLP:journals/tomacs/AghaP18}
Agha, G., Palmskog, K.: A survey of statistical model checking.
\newblock {ACM} Trans. Model. Comput. Simul. \textbf{28}(1), 6:1--6:39 (2018)

\bibitem{DBLP:conf/lics/Alfaro98}
de~Alfaro, L.: How to specify and verify the long-run average behavior of
  probabilistic systems.
\newblock In: {LICS}, pp. 454--465. {IEEE} Computer Society (1998)

\bibitem{DBLP:journals/siglog/AlurHV15}
Alur, R., Henzinger, T.A., Vardi, M.Y.: Theory in practice for system design
  and verification.
\newblock {SIGLOG} News \textbf{2}(1), 46--51 (2015)

\bibitem{DBLP:journals/aamas/AmatoBZ10}
Amato, C., Bernstein, D.S., Zilberstein, S.: Optimizing fixed-size stochastic
  controllers for {POMDP}s and decentralized {POMDP}s.
\newblock Autonomous Agents and Multi-Agent Systems \textbf{21}(3), 293--320
  (2010)

\bibitem{DBLP:conf/formats/AndovaHK03}
Andova, S., Hermanns, H., Katoen, J.P.: Discrete-time rewards model-checked.
\newblock In: {FORMATS}, \emph{{LNCS}}, vol. 2791, pp. 88--104. Springer (2003)

\bibitem{DBLP:conf/cav/AshokCDKM17}
Ashok, P., Chatterjee, K., Daca, P., Kret{\'{\i}}nsk{\'{y}}, J., Meggendorfer,
  T.: Value iteration for long-run average reward in {M}arkov decision
  processes.
\newblock In: {CAV} {(1)}, \emph{{LNCS}}, vol. 10426, pp. 201--221. Springer
  (2017)

\bibitem{Astrom65}
{\AA}str{\"o}m, K.: Optimal control of {M}arkov processes with incomplete state
  information.
\newblock Journal of Mathematical Analysis and Applications \textbf{10}(1), 174
  -- 205 (1965)

\bibitem{DBLP:journals/tocl/AzizSSB00}
Aziz, A., Sanwal, K., Singhal, V., Brayton, R.K.: Model-checking continous-time
  {M}arkov chains.
\newblock {ACM} Trans. Comput. Log. \textbf{1}(1), 162--170 (2000)

\bibitem{DBLP:reference/mc/BaierAFK18}
Baier, C., de~Alfaro, L., Forejt, V., Kwiatkowska, M.: Model checking
  probabilistic systems.
\newblock In: Handbook of Model Checking, pp. 963--999. Springer (2018)

\bibitem{DBLP:conf/icalp/BaierCHKR97}
Baier, C., Clarke, E.M., Hartonas{-}Garmhausen, V., Kwiatkowska, M.Z., Ryan,
  M.: Symbolic model checking for probabilistic processes.
\newblock In: {ICALP}, \emph{{LNCS}}, vol. 1256, pp. 430--440. Springer (1997)

\bibitem{DBLP:journals/tse/BaierHHK03}
Baier, C., Haverkort, B.R., Hermanns, H., Katoen, J.: Model-checking algorithms
  for continuous-time {M}arkov chains.
\newblock {IEEE} Trans. Software Eng. \textbf{29}(6), 524--541 (2003)

\bibitem{DBLP:books/daglib/0020348}
Baier, C., Katoen, J.P.: Principles of Model Checking.
\newblock {MIT} Press (2008)

\bibitem{DBLP:conf/tacas/BaierKKM14}
Baier, C., Klein, J., Kl{\"{u}}ppelholz, S., M{\"{a}}rcker, S.: Computing
  conditional probabilities in {M}arkovian models efficiently.
\newblock In: {TACAS}, \emph{{LNCS}}, vol. 8413, pp. 515--530. Springer (2014)

\bibitem{DBLP:conf/tacas/Baier0KW17}
Baier, C., Klein, J., Kl{\"{u}}ppelholz, S., Wunderlich, S.: Maximizing the
  conditional expected reward for reaching the goal.
\newblock In: {TACAS} {(2)}, \emph{{LNCS}}, vol. 10206, pp. 269--285 (2017)

\bibitem{DBLP:conf/cav/Baier0L0W17}
Baier, C., Klein, J., Leuschner, L., Parker, D., Wunderlich, S.: Ensuring the
  reliability of your model checker: Interval iteration for {M}arkov decision
  processes.
\newblock In: {CAV} {(1)}, \emph{{LNCS}}, vol. 10426, pp. 160--180. Springer
  (2017)

\bibitem{DBLP:journals/cacm/BallLR11}
Ball, T., Levin, V., Rajamani, S.K.: A decade of software model checking with
  {SLAM}.
\newblock Commun. {ACM} \textbf{54}(7), 68--76 (2011)

\bibitem{BFT15}
Barrett, C., Fontaine, P., Tinelli, C.: The {SMT-LIB} standard: Version 2.5.
\newblock Tech. rep., Dep. of Computer Science,The University of Iowa (2015).
\newblock \href{http://www.smt-lib.org/}{www.smt-lib.org}

\bibitem{DBLP:conf/fmcad/BauerMCS017}
Bauer, M.S., Mathur, U., Chadha, R., Sistla, A.P., Viswanathan, M.: Exact
  quantitative probabilistic model checking through rational search.
\newblock In: {FMCAD}, pp. 92--99. {IEEE} (2017)

\bibitem{DBLP:journals/corr/abs-2007-00102}
Bork, A., Junges, S., Katoen, J., Quatmann, T.: Verification of
  indefinite-horizon {POMDPs}.
\newblock CoRR \textbf{abs/2007.00102} (2020)

\bibitem{DBLP:conf/atva/BoudaliCS07}
Boudali, H., Crouzen, P., Stoelinga, M.: A compositional semantics for dynamic
  fault trees in terms of interactive {M}arkov chains.
\newblock In: {ATVA}, \emph{{LNCS}}, vol. 4762, pp. 441--456. Springer (2007)

\bibitem{DBLP:conf/dsn/BoudaliCS07}
Boudali, H., Crouzen, P., Stoelinga, M.: Dynamic fault tree analysis using
  input/output interactive {M}arkov chains.
\newblock In: {DSN}, pp. 708--717. {IEEE} Computer Society (2007)

\bibitem{DBLP:journals/cj/BozzanoCKNNR11}
Bozzano, M., Cimatti, A., Katoen, J.P., Nguyen, V.Y., Noll, T., Roveri, M.:
  Safety, dependability and performance analysis of extended {AADL} models.
\newblock Comput. J. \textbf{54}(5), 754--775 (2011)

\bibitem{DBLP:conf/atva/BrazdilCCFKKPU14}
Br{\'{a}}zdil, T., Chatterjee, K., Chmelik, M., Forejt, V.,
  Kret{\'{\i}}nsk{\'{y}}, J., Kwiatkowska, M.Z., Parker, D., Ujma, M.:
  Verification of {M}arkov decision processes using learning algorithms.
\newblock In: {ATVA}, \emph{{LNCS}}, vol. 8837, pp. 98--114. Springer (2014)

\bibitem{DBLP:conf/aaai/BraziunasB04}
Braziunas, D., Boutilier, C.: Stochastic local search for {{POMDP}}
  controllers.
\newblock In: {AAAI}, pp. 690--696. {AAAI} Press / The {MIT} Press (2004)

\bibitem{DBLP:conf/tacas/BuddeDHHJT17}
Budde, C.E., Dehnert, C., Hahn, E.M., Hartmanns, A., Junges, S., Turrini, A.:
  {JANI:} quantitative model and tool interaction.
\newblock In: {TACAS} {(2)}, \emph{{LNCS}}, vol. 10206, pp. 151--168 (2017)

\bibitem{qcomp2020}
Budde, C.E., Hartmanns, A., Klauck, M., Kret{\'{\i}}nsk{\'{y}}, J., Parker, D.,
  Quatmann, T., Turini, A., Zhang, Z.: On correctness, precision, and
  performance in quantitative verification ({QC}omp 2020 competition report).
\newblock In: {ISoLA}, {LNCS}. Springer (2020).
\newblock To Appear.

\bibitem{DBLP:conf/qest/ButkovaHH19}
Butkova, Y., Hartmanns, A., Hermanns, H.: A {M}odest approach to modelling and
  checking {M}arkov automata.
\newblock In: {QEST}, \emph{{LNCS}}, vol. 11785, pp. 52--69. Springer (2019)

\bibitem{DBLP:conf/tacas/ButkovaWH17}
Butkova, Y., Wimmer, R., Hermanns, H.: Long-run rewards for {M}arkov automata.
\newblock In: {TACAS} {(2)}, \emph{{LNCS}}, vol. 10206, pp. 188--203 (2017)

\bibitem{DBLP:journals/tcsb/CalderVGO06}
Calder, M., Vyshemirsky, V., Gilbert, D.R., Orton, R.J.: Analysis of signalling
  pathways using continuous time {M}arkov chains.
\newblock In: Transactions on Computational Systems Biology {VI},
  \emph{{LNCS}}, vol. 4220, pp. 44--67 (2006)

\bibitem{DBLP:conf/fm/CeskaHJK19}
Ceska, M., Hensel, C., Junges, S., Katoen, J.P.: Counterexample-driven
  synthesis for probabilistic program sketches.
\newblock In: {FM}, \emph{{LNCS}}, vol. 11800, pp. 101--120. Springer (2019)

\bibitem{DBLP:journals/tocl/ChadhaV10}
Chadha, R., Viswanathan, M.: A counterexample-guided abstraction-refinement
  framework for {M}arkov decision processes.
\newblock {ACM} Trans. Comput. Log. \textbf{12}(1), 1:1--1:49 (2010)

\bibitem{DBLP:conf/aaai/ChatterjeeCD16}
Chatterjee, K., Chmelik, M., Davies, J.: A symbolic {SAT}-based algorithm for
  almost-sure reachability with small strategies in {POMDPs}.
\newblock In: {AAAI}, pp. 3225--3232. {AAAI} Press (2016)

\bibitem{DBLP:conf/mfcs/ChatterjeeDH10}
Chatterjee, K., Doyen, L., Henzinger, T.A.: Qualitative analysis of
  partially-observable {M}arkov decision processes.
\newblock In: {MFCS}, \emph{{LNCS}}, vol. 6281, pp. 258--269. Springer (2010)

\bibitem{DBLP:conf/tacas/CimattiGSS13}
Cimatti, A., Griggio, A., Schaafsma, B.J., Sebastiani, R.: The mathsat5 {SMT}
  solver.
\newblock In: {TACAS}, \emph{{LNCS}}, vol. 7795, pp. 93--107. Springer (2013)

\bibitem{DBLP:conf/dimacs/Condon90}
Condon, A.: On algorithms for simple stochastic games.
\newblock In: Advances In Computational Complexity Theory, \emph{{DIMACS}
  Series in Discrete Mathematics and Theoretical Computer Science}, vol.~13,
  pp. 51--71. {DIMACS/AMS} (1990)

\bibitem{DBLP:conf/sat/CorziliusKJSA15}
Corzilius, F., Kremer, G., Junges, S., Schupp, S., {\'{A}}brah{\'{a}}m, E.:
  {SMT-RAT:} an open source {C++} toolbox for strategic and parallel {SMT}
  solving.
\newblock In: {SAT}, \emph{{LNCS}}, vol. 9340, pp. 360--368. Springer (2015)

\bibitem{DBLP:conf/focs/CourcoubetisY88}
Courcoubetis, C., Yannakakis, M.: Verifying temporal properties of finite-state
  probabilistic programs.
\newblock In: {FOCS}, pp. 338--345. {IEEE} Computer Society (1988)

\bibitem{DBLP:conf/ictac/Daws04}
Daws, C.: Symbolic and parametric model checking of discrete-time {M}arkov
  chains.
\newblock In: {ICTAC}, \emph{{LNCS}}, vol. 3407, pp. 280--294. Springer (2004)

\bibitem{DBLP:conf/atva/DehnertJWAK14}
Dehnert, C., Jansen, N., Wimmer, R., {\'{A}}brah{\'{a}}m, E., Katoen, J.P.:
  Fast debugging of {PRISM} models.
\newblock In: {ATVA}, \emph{{LNCS}}, vol. 8837, pp. 146--162. Springer (2014)

\bibitem{DBLP:conf/cav/DehnertJJCVBKA15}
Dehnert, C., Junges, S., Jansen, N., Corzilius, F., Volk, M., Bruintjes, H.,
  Katoen, J.P., {\'{A}}brah{\'{a}}m, E.: Prophesy: {A} probabilistic parameter
  synthesis tool.
\newblock In: {CAV} {(1)}, \emph{{LNCS}}, vol. 9206, pp. 214--231. Springer
  (2015)

\bibitem{DBLP:conf/cav/DehnertJK017}
Dehnert, C., Junges, S., Katoen, J.P., Volk, M.: A storm is coming: {A} modern
  probabilistic model checker.
\newblock In: {CAV} {(2)}, \emph{{LNCS}}, vol. 10427, pp. 592--600. Springer
  (2017)

\bibitem{DBLP:conf/vmcai/DehnertKP13}
Dehnert, C., Katoen, J.P., Parker, D.: {SMT}-based bisimulation minimisation of
  {M}arkov models.
\newblock In: {VMCAI}, \emph{{LNCS}}, vol. 7737, pp. 28--47. Springer (2013)

\bibitem{DBLP:conf/tacas/DelgrangeKQR20}
Delgrange, F., Katoen, J., Quatmann, T., Randour, M.: Simple strategies in
  multi-objective {MDP}s.
\newblock In: {TACAS} {(1)}, \emph{LNCS}, vol. 12078, pp. 346--364. Springer
  (2020)

\bibitem{DBLP:phd/basesearch/Dijk16}
van Dijk, T.: Sylvan: multi-core decision diagrams.
\newblock Ph.D. thesis, University of Twente, Enschede, Netherlands (2016)

\bibitem{DBLP:journals/sttt/DijkP18}
van Dijk, T., van~de Pol, J.: Multi-core symbolic bisimulation minimisation.
\newblock {STTT} \textbf{20}(2), 157--177 (2018)

\bibitem{DBLP:journals/corr/DragerFK0U15}
Dr{\"{a}}ger, K., Forejt, V., Kwiatkowska, M.Z., Parker, D., Ujma, M.:
  Permissive controller synthesis for probabilistic systems.
\newblock Logical Methods in Computer Science \textbf{11}(2) (2015)

\bibitem{DBB90}
Dugan, J.B., Bavuso, S.J., Boyd, M.: Fault trees and sequence dependencies.
\newblock In: Proc.\ of RAMS, pp. 286--293. IEEE (1990).
\newblock \doi{10.1109/ARMS.1990.67971}

\bibitem{DBLP:conf/apn/EisentrautHK013}
Eisentraut, C., Hermanns, H., Katoen, J.P., Zhang, L.: A semantics for every
  {GSPN}.
\newblock In: Petri Nets, \emph{{LNCS}}, vol. 7927, pp. 90--109. Springer
  (2013)

\bibitem{DBLP:conf/lics/EisentrautHZ10}
Eisentraut, C., Hermanns, H., Zhang, L.: On probabilistic automata in
  continuous time.
\newblock In: {LICS}, pp. 342--351. {IEEE} Computer Society (2010)

\bibitem{DBLP:journals/lmcs/EtessamiKVY08}
Etessami, K., Kwiatkowska, M.Z., Vardi, M.Y., Yannakakis, M.: Multi-objective
  model checking of {M}arkov decision processes.
\newblock Logical Methods in Computer Science \textbf{4}(4) (2008)

\bibitem{DBLP:conf/tacas/ForejtKNPQ11}
Forejt, V., Kwiatkowska, M.Z., Norman, G., Parker, D., Qu, H.: Quantitative
  multi-objective verification for probabilistic systems.
\newblock In: {TACAS}, \emph{{LNCS}}, vol. 6605, pp. 112--127. Springer (2011)

\bibitem{DBLP:conf/atva/ForejtKP12}
Forejt, V., Kwiatkowska, M.Z., Parker, D.: {P}areto curves for probabilistic
  model checking.
\newblock In: {ATVA}, \emph{{LNCS}}, vol. 7561, pp. 317--332. Springer (2012)

\bibitem{Fredlund1994TheTA}
Fredlund, L.: The timing and probability workbench: A tool for analysing timed
  processes.
\newblock Tech. Rep.~49, Uppsala University (1994)

\bibitem{DBLP:journals/ress/GhadhabJKKV19}
Ghadhab, M., Junges, S., Katoen, J.P., Kuntz, M., Volk, M.: Safety analysis for
  vehicle guidance systems with dynamic fault trees.
\newblock Rel. Eng. {\&} Sys. Safety \textbf{186}, 37--50 (2019)

\bibitem{DBLP:conf/icse/GordonHNR14}
Gordon, A.D., Henzinger, T.A., Nori, A.V., Rajamani, S.K.: Probabilistic
  programming.
\newblock In: {FOSE}, pp. 167--181. {ACM} (2014)

\bibitem{eigenweb}
Guennebaud, G., Jacob, B., et~al.: Eigen v3.
\newblock http://eigen.tuxfamily.org (2010)

\bibitem{gurobi_website}
Gurobi~Optimization, L.: Gurobi optimizer reference manual (2019).
\newblock \urlprefix\url{http://www.gurobi.com}

\bibitem{DBLP:conf/rp/HaddadM14}
Haddad, S., Monmege, B.: Reachability in {MDP}s: Refining convergence of value
  iteration.
\newblock In: {RP}, \emph{{LNCS}}, vol. 8762, pp. 125--137. Springer (2014)

\bibitem{DBLP:conf/setta/HahnH16}
Hahn, E.M., Hartmanns, A.: A comparison of time- and reward-bounded
  probabilistic model checking techniques.
\newblock In: {SETTA}, \emph{{LNCS}}, vol. 9984, pp. 85--100 (2016)

\bibitem{DBLP:conf/tacas/HahnHHKKKPQRS19}
Hahn, E.M., Hartmanns, A., Hensel, C., Klauck, M., Klein, J.,
  Kret{\'{\i}}nsk{\'{y}}, J., Parker, D., Quatmann, T., Ruijters, E.,
  Steinmetz, M.: The 2019 comparison of tools for the analysis of quantitative
  formal models - ({QComp} 2019 competition report).
\newblock In: {TACAS} {(3)}, \emph{{LNCS}}, vol. 11429, pp. 69--92. Springer
  (2019)

\bibitem{DBLP:journals/sttt/HahnHZ11}
Hahn, E.M., Hermanns, H., Zhang, L.: Probabilistic reachability for parametric
  {M}arkov models.
\newblock {STTT} \textbf{13}(1), 3--19 (2011)

\bibitem{DBLP:conf/fm/HahnLSTZ14}
Hahn, E.M., Li, Y., Schewe, S., Turrini, A., Zhang, L.: iscas{M}c: {A}
  web-based probabilistic model checker.
\newblock In: {FM}, \emph{{LNCS}}, vol. 8442, pp. 312--317. Springer (2014)

\bibitem{DBLP:journals/tse/HanKD09}
Han, T., Katoen, J.P., Damman, B.: Counterexample generation in probabilistic
  model checking.
\newblock {IEEE} Trans. Software Eng. \textbf{35}(2), 241--257 (2009)

\bibitem{DBLP:conf/uai/Hansen98}
Hansen, E.A.: Solving {POMDP}s by searching in policy space.
\newblock In: {UAI}, pp. 211--219. Morgan Kaufmann (1998)

\bibitem{DBLP:conf/rtss/HanssonJ89}
Hansson, H., Jonsson, B.: A framework for reasoning about time and reliability.
\newblock In: {RTSS}, pp. 102--111. {IEEE} Computer Society (1989)

\bibitem{DBLP:journals/fac/HanssonJ94}
Hansson, H., Jonsson, B.: A logic for reasoning about time and reliability.
\newblock Formal Asp. Comput. \textbf{6}(5), 512--535 (1994)

\bibitem{DBLP:conf/tacas/HartmannsH14}
Hartmanns, A., Hermanns, H.: The {M}odest {T}oolset: An integrated environment
  for quantitative modelling and verification.
\newblock In: {TACAS}, \emph{{LNCS}}, vol. 8413, pp. 593--598. Springer (2014)

\bibitem{DBLP:conf/atva/HartmannsH15}
Hartmanns, A., Hermanns, H.: Explicit model checking of very large {MDP} using
  partitioning and secondary storage.
\newblock In: {ATVA}, \emph{{LNCS}}, vol. 9364, pp. 131--147. Springer (2015)

\bibitem{DBLP:conf/tacas/HartmannsJKQ18}
Hartmanns, A., Junges, S., Katoen, J.P., Quatmann, T.: Multi-cost bounded
  reachability in {MDP}.
\newblock In: {TACAS} {(2)}, \emph{{LNCS}}, vol. 10806, pp. 320--339. Springer
  (2018)

\bibitem{HJKQ19}
Hartmanns, A., Junges, S., Katoen, J.P., Quatmann, T.: Multi-cost bounded
  tradeoff analysis in {MDP}.
\newblock JAR  (2020)

\bibitem{DBLP:conf/cav/HartmannsK20}
Hartmanns, A., Kaminski, B.L.: Optimistic value iteration.
\newblock In: {CAV} {(2)}, \emph{LNCS}, vol. 12225, pp. 488--511. Springer
  (2020)

\bibitem{DBLP:conf/tacas/HartmannsKPQR19}
Hartmanns, A., Klauck, M., Parker, D., Quatmann, T., Ruijters, E.: The
  quantitative verification benchmark set.
\newblock In: {TACAS} {(1)}, \emph{{LNCS}}, vol. 11427, pp. 344--350. Springer
  (2019)

\bibitem{DBLP:conf/arts/Hartonas-GarmhausenCC99}
Hartonas{-}Garmhausen, V., Campos, S.V.A., Clarke, E.M.: Prob{V}erus:
  Probabilistic symbolic model checking.
\newblock In: {ARTS}, \emph{{LNCS}}, vol. 1601, pp. 96--110. Springer (1999)

\bibitem{DBLP:journals/scp/JifengSM97}
He, J., Seidel, K., McIver, A.: Probabilistic models for the guarded command
  language.
\newblock Sci. Comput. Program. \textbf{28}(2-3), 171--192 (1997)

\bibitem{DBLP:conf/types/HelminkSV93}
Helmink, L., Sellink, M.P.A., Vaandrager, F.W.: Proof-checking a data link
  protocol.
\newblock In: {TYPES}, \emph{{LNCS}}, vol. 806, pp. 127--165. Springer (1993)

\bibitem{DBLP:phd/dnb/Hensel18}
Hensel, C.: The probabilistic model checker {S}torm: symbolic methods for
  probabilistic model checking.
\newblock Ph.D. thesis, {RWTH} Aachen University, Germany (2018)

\bibitem{replication_package}
Hensel, C., Junges, S., Katoen, J.P., Quatmann, T., Volk, M.: {The
  Probabilistic Model Checker {S}torm: Evaluation Results and Replication
  Package} (2020).
\newblock \urlprefix\url{https://doi.org/10.5281/zenodo.3571209}

\bibitem{DBLP:conf/tacas/HermannsKMS00}
Hermanns, H., Katoen, J.P., Meyer{-}Kayser, J., Siegle, M.: A {M}arkov chain
  model checker.
\newblock In: {TACAS}, \emph{{LNCS}}, vol. 1785, pp. 347--362. Springer (2000)

\bibitem{DBLP:journals/cacm/Holzmann14}
Holzmann, G.J.: Mars code.
\newblock Commun. {ACM} \textbf{57}(2), 64--73 (2014)

\bibitem{DBLP:conf/ijcai/HorakBC18}
Hor{\'{a}}k, K., Bosansk{\'{y}}, B., Chatterjee, K.: Goal-{HSVI}: Heuristic
  search value iteration for goal {POMDP}s.
\newblock In: {IJCAI}, pp. 4764--4770. ijcai.org (2018)

\bibitem{DBLP:journals/corr/abs-1903-07993}
Junges, S., {\'{A}}brah{\'{a}}m, E., Hensel, C., Jansen, N., Katoen, J.P.,
  Quatmann, T., Volk, M.: Parameter synthesis for {M}arkov models.
\newblock CoRR \textbf{abs/1903.07993} (2019)

\bibitem{DBLP:conf/tacas/Junges0DTK16}
Junges, S., Jansen, N., Dehnert, C., Topcu, U., Katoen, J.P.:
  Safety-constrained reinforcement learning for mdps.
\newblock In: {TACAS}, \emph{{LNCS}}, vol. 9636, pp. 130--146. Springer (2016)

\bibitem{DBLP:journals/corr/abs-2007-00085}
Junges, S., Jansen, N., Seshia, S.A.: Enforcing almost-sure reachability in
  pomdps.
\newblock CoRR \textbf{abs/2007.00085} (2020)

\bibitem{DBLP:conf/uai/Junges0WQWK018}
Junges, S., Jansen, N., Wimmer, R., Quatmann, T., Winterer, L., Katoen, J.P.,
  Becker, B.: Finite-state controllers of {POMDP}s using parameter synthesis.
\newblock In: {UAI}, pp. 519--529. {AUAI} Press (2018)

\bibitem{DBLP:journals/ai/KaelblingLC98}
Kaelbling, L.P., Littman, M.L., Cassandra, A.R.: Planning and acting in
  partially observable stochastic domains.
\newblock Artif. Intell. \textbf{101}(1-2), 99--134 (1998)

\bibitem{DBLP:conf/lics/Katoen16}
Katoen, J.P.: The probabilistic model checking landscape.
\newblock In: {LICS}, pp. 31--45. {ACM} (2016)

\bibitem{DBLP:conf/tacas/KatoenKZJ07}
Katoen, J.P., Kemna, T., Zapreev, I.S., Jansen, D.N.: Bisimulation minimisation
  mostly speeds up probabilistic model checking.
\newblock In: {TACAS}, \emph{{LNCS}}, vol. 4424, pp. 87--101. Springer (2007)

\bibitem{DBLP:journals/pe/KatoenZHHJ11}
Katoen, J.P., Zapreev, I.S., Hahn, E.M., Hermanns, H., Jansen, D.N.: The ins
  and outs of the probabilistic model checker {MRMC}.
\newblock Perform. Eval. \textbf{68}(2), 90--104 (2011)

\bibitem{DBLP:journals/sttt/KleinBCDDKMM18}
Klein, J., Baier, C., Chrszon, P., Daum, M., Dubslaff, C., Kl{\"{u}}ppelholz,
  S., M{\"{a}}rcker, S., M{\"{u}}ller, D.: Advances in probabilistic model
  checking with {PRISM:} variable reordering, quantiles and weak deterministic
  b{\"{u}}chi automata.
\newblock {STTT} \textbf{20}(2), 179--194 (2018)

\bibitem{DBLP:journals/ipl/KwekM03}
Kwek, S., Mehlhorn, K.: Optimal search for rationals.
\newblock Inf. Process. Lett. \textbf{86}(1), 23--26 (2003)

\bibitem{DBLP:conf/tacas/KwiatkowskaNP02}
Kwiatkowska, M.Z., Norman, G., Parker, D.: Probabilistic symbolic model
  checking with {PRISM:} {A} hybrid approach.
\newblock In: {TACAS}, \emph{{LNCS}}, vol. 2280, pp. 52--66. Springer (2002)

\bibitem{DBLP:conf/qest/KwiatkowskaNP06}
Kwiatkowska, M.Z., Norman, G., Parker, D.: Game-based abstraction for {M}arkov
  decision processes.
\newblock In: {QEST}, pp. 157--166. {IEEE} Computer Society (2006)

\bibitem{DBLP:conf/cav/KwiatkowskaNP11}
Kwiatkowska, M.Z., Norman, G., Parker, D.: {PRISM} 4.0: Verification of
  probabilistic real-time systems.
\newblock In: {CAV}, \emph{{LNCS}}, vol. 6806, pp. 585--591. Springer (2011)

\bibitem{DBLP:journals/fac/KwiatkowskaNP12}
Kwiatkowska, M.Z., Norman, G., Parker, D.: Probabilistic verification of
  herman's self-stabilisation algorithm.
\newblock Formal Asp. Comput. \textbf{24}(4-6), 661--670 (2012)

\bibitem{DBLP:conf/cav/KwiatkowskaNS01}
Kwiatkowska, M.Z., Norman, G., Segala, R.: Automated verification of a
  randomized distributed consensus protocol using cadence {SMV} and {PRISM}.
\newblock In: {CAV}, \emph{{LNCS}}, vol. 2102, pp. 194--206. Springer (2001)

\bibitem{DBLP:journals/fac/LanotteMT07}
Lanotte, R., Maggiolo{-}Schettini, A., Troina, A.: Parametric probabilistic
  transition systems for system design and analysis.
\newblock Formal Asp. Comput. \textbf{19}(1), 93--109 (2007)

\bibitem{DBLP:conf/isola/LarsenL16}
Larsen, K.G., Legay, A.: Statistical model checking: Past, present, and future.
\newblock In: ISoLA {(1)}, \emph{{LNCS}}, vol. 9952, pp. 3--15 (2016)

\bibitem{DBLP:journals/ior/Lovejoy91}
Lovejoy, W.S.: Computationally feasible bounds for partially observed {M}arkov
  decision processes.
\newblock Oper. Res. \textbf{39}(1), 162--175 (1991)

\bibitem{DBLP:journals/ai/MadaniHC03}
Madani, O., Hanks, S., Condon, A.: On the undecidability of probabilistic
  planning and related stochastic optimization problems.
\newblock Artif. Intell. \textbf{147}(1-2), 5--34 (2003)

\bibitem{DBLP:journals/tocs/MarsanCB84}
Marsan, M.A., Conte, G., Balbo, G.: A class of generalized stochastic petri
  nets for the performance evaluation of multiprocessor systems.
\newblock {ACM} Trans. Comput. Syst. \textbf{2}(2), 93--122 (1984)

\bibitem{DBLP:conf/uai/MeuleauKKC99}
Meuleau, N., Kim, K., Kaelbling, L.P., Cassandra, A.R.: Solving {POMDP}s by
  searching the space of finite policies.
\newblock In: {UAI}, pp. 417--426. Morgan Kaufmann (1999)

\bibitem{DBLP:conf/tacas/MouraB08}
de~Moura, L.M., Bj{\o}rner, N.: {Z3:} an efficient {SMT} solver.
\newblock In: {TACAS}, \emph{{LNCS}}, vol. 4963, pp. 337--340. Springer (2008)

\bibitem{DBLP:journals/rts/Norman0Z17}
Norman, G., Parker, D., Zou, X.: Verification and control of partially
  observable probabilistic systems.
\newblock Real-Time Systems \textbf{53}(3), 354--402 (2017)

\bibitem{DBLP:books/daglib/0095301}
Norris, J.R.: {M}arkov chains.
\newblock Cambridge series in statistical and probabilistic mathematics.
  Cambridge University Press (1998)

\bibitem{DBLP:journals/toplas/OlmedoGJKKM18}
Olmedo, F., Gretz, F., Jansen, N., Kaminski, B.L., Katoen, J.P., McIver, A.:
  Conditioning in probabilistic programming.
\newblock {ACM} Trans. Program. Lang. Syst. \textbf{40}(1), 4:1--4:50 (2018)

\bibitem{DBLP:conf/nips/PajarinenP11}
Pajarinen, J., Peltonen, J.: Periodic finite state controllers for efficient
  {{POMDP}} and {DEC-{POMDP}} planning.
\newblock In: {NIPS}, pp. 2636--2644 (2011)

\bibitem{scikit}
Pedregosa, F., Varoquaux, G., Gramfort, A., Michel, V., Thirion, B., Grisel,
  O., Blondel, M., Prettenhofer, P., Weiss, R., Dubourg, V., VanderPlas, J.,
  Passos, A., Cournapeau, D., Brucher, M., Perrot, M., Duchesnay, E.:
  Scikit-learn: Machine learning in python.
\newblock J. Mach. Learn. Res. \textbf{12}, 2825--2830 (2011)

\bibitem{Put94}
Puterman, M.L.: {{M}arkov} Decision Processes.
\newblock John Wiley and Sons (1994)

\bibitem{DBLP:conf/atva/QuatmannD0JK16}
Quatmann, T., Dehnert, C., Jansen, N., Junges, S., Katoen, J.P.: Parameter
  synthesis for {M}arkov models: Faster than ever.
\newblock In: {ATVA}, \emph{{LNCS}}, vol. 9938, pp. 50--67 (2016)

\bibitem{DBLP:conf/cav/QuatmannJK17}
Quatmann, T., Junges, S., Katoen, J.P.: {M}arkov automata with multiple
  objectives.
\newblock In: {CAV} {(1)}, \emph{{LNCS}}, vol. 10426, pp. 140--159. Springer
  (2017)

\bibitem{DBLP:conf/cav/QuatmannK18}
Quatmann, T., Katoen, J.P.: Sound value iteration.
\newblock In: {CAV} {(1)}, \emph{{LNCS}}, vol. 10981, pp. 643--661. Springer
  (2018)

\bibitem{DBLP:journals/csr/RuijtersS15}
Ruijters, E., Stoelinga, M.: Fault tree analysis: {A} survey of the
  state-of-the-art in modeling, analysis and tools.
\newblock Computer Science Review \textbf{15}, 29--62 (2015)

\bibitem{DBLP:journals/njc/SegalaL95}
Segala, R., Lynch, N.A.: Probabilistic simulations for probabilistic processes.
\newblock Nord. J. Comput. \textbf{2}(2), 250--273 (1995)

\bibitem{cudd_website}
Somenzi, F.: \cudd{} 3.0.0.
\newblock \urlprefix\url{http://vlsi.colorado.edu/\textasciitilde
  fabio/CUDD/html/}.
\newblock Also available at \url{https://github.com/ivmai/cudd}

\bibitem{DBLP:conf/atva/SpelJK19}
Spel, J., Junges, S., Katoen, J.P.: Are parametric {M}arkov chains monotonic?
\newblock In: {ATVA}, \emph{{LNCS}}, vol. 11781, pp. 479--496. Springer (2019)

\bibitem{DBLP:conf/ftcs/SullivanDC99}
Sullivan, K.J., Dugan, J.B., Coppit, D.: The galileo fault tree analysis tool.
\newblock In: {FTCS}, pp. 232--235. {IEEE} Computer Society (1999)

\bibitem{DBLP:conf/focs/Vardi85}
Vardi, M.Y.: Automatic verification of probabilistic concurrent finite-state
  programs.
\newblock In: {FOCS}, pp. 327--338. {IEEE} Computer Society (1985)

\bibitem{DBLP:journals/tii/VolkJK18}
Volk, M., Junges, S., Katoen, J.P.: Fast dynamic fault tree analysis by model
  checking techniques.
\newblock {IEEE} Trans. Industrial Informatics \textbf{14}(1), 370--379 (2018)

\bibitem{DBLP:phd/de/Wachter2011}
Wachter, B.: Refined {P}robabilistic {A}bstraction.
\newblock Ph.D. thesis, Saarland University (2011)

\bibitem{DBLP:conf/gi/Wimmer11}
Wimmer, R.: Symbolische {M}ethoden f{\"{u}}r die probabilistische
  {V}erifikation: {Z}ustandsraumreduktion und {G}egenbeispiele.
\newblock In: Ausgezeichnete Informatikdissertationen, \emph{{LNI}}, vol.
  {D-12}, pp. 271--280. {GI} (2011)

\bibitem{DBLP:conf/qest/WimmerJVAKB13}
Wimmer, R., Jansen, N., Vorpahl, A., {\'{A}}brah{\'{a}}m, E., Katoen, J.P.,
  Becker, B.: High-level counterexamples for probabilistic automata.
\newblock In: {QEST}, \emph{{LNCS}}, vol. 8054, pp. 39--54. Springer (2013)

\bibitem{DBLP:conf/ddecs/WimmerKHB08}
Wimmer, R., Kortus, A., Herbstritt, M., Becker, B.: Probabilistic model
  checking and reliability of results.
\newblock In: {DDECS}, pp. 207--212. {IEEE} Computer Society (2008)

\bibitem{DBLP:conf/concur/WinklerJPK19}
Winkler, T., Junges, S., P{\'{e}}rez, G.A., Katoen, J.: On the complexity of
  reachability in parametric markov decision processes.
\newblock In: {CONCUR}, \emph{LIPIcs}, vol. 140, pp. 14:1--14:17. Schloss
  Dagstuhl - Leibniz-Zentrum f{\"{u}}r Informatik (2019)

\bibitem{DBLP:conf/cdc/WintererJW0TK017}
Winterer, L., Junges, S., Wimmer, R., Jansen, N., Topcu, U., Katoen, J.P.,
  Becker, B.: Motion planning under partial observability using game-based
  abstraction.
\newblock In: {CDC}, pp. 2201--2208. {IEEE} (2017)

\end{thebibliography}
